\documentclass[sigconf]{acmart}

\AtBeginDocument{%
  \providecommand\BibTeX{{%
    \normalfont B\kern-0.5em{\scshape i\kern-0.25em b}\kern-0.8em\TeX}}}

\usepackage{amsmath}
\usepackage{graphics}
\usepackage{siunitx}
\usepackage[title]{appendix}
\usepackage{enumitem}
\usepackage{balance}
\usepackage{xspace}
\usepackage{color, colortbl, soul, xcolor}
\usepackage{subcaption}
\usepackage{fixltx2e}
\usepackage{bm}
\usepackage{mathtools}
\usepackage{algorithm}
\usepackage{makecell}
\usepackage{algorithmic}

\newcommand\algorithmicprocedure{\textbf{function}}
\newcommand{\algorithmicendprocedure}{\algorithmicend\ \algorithmicprocedure}
\makeatletter
\newcommand\PROCEDURE[3][default]{%
  \ALC@it
  \algorithmicprocedure\ \textsc{#2}(#3)%
  \ALC@com{#1}%
  \begin{ALC@prc}%
}
\newcommand\ENDPROCEDURE{%
  \end{ALC@prc}%
  \ifthenelse{\boolean{ALC@noend}}{}{%
    \ALC@it\algorithmicendprocedure
  }%
}
\newenvironment{ALC@prc}{\begin{ALC@g}}{\end{ALC@g}}
\makeatother

\def\S{Sec.\xspace}

\def\insitu{\textit{in situ}\xspace}
\def\ie{\textit{i.e.,}\xspace}
\def\etal{\textit{et~al.}\xspace}
\def\etc{\textit{etc.}\xspace}
\def\eg{\textit{e.g.,}\xspace}
\def\incl{\textit{incl.}\xspace}

\def\first{\textit{first}\xspace}

\def\second{\textit{second}\xspace}

\def\posthoc{\textit{post-hoc}\xspace}
\def\insitu{\textit{in-situ}\xspace}

\def\sysname{PaperToPlace}

\newcommand{\red}[1]{\textcolor{red}{#1}}
\newcommand{\green}[1]{\textcolor{green}{#1}}
\newcommand{\blue}[1]{\textcolor{blue}{#1}}

\DeclareMathOperator*{\argmin}{argmin}

\copyrightyear{2023} 
\acmYear{2023} 
\setcopyright{rightsretained} 
\acmConference[UIST '23]{The 36th Annual ACM Symposium on User Interface Software and Technology}{October 29-November 1, 2023}{San Francisco, CA, USA}
\acmBooktitle{The 36th Annual ACM Symposium on User Interface Software and Technology (UIST '23), October 29-November 1, 2023, San Francisco, CA, USA}
\acmDOI{10.1145/3586183.3606832}
\acmISBN{979-8-4007-0132-0/23/10}

\ccsdesc[500]{Human-centered computing~Mixed / augmented reality}

\begin{teaserfigure}
    \includegraphics[width=\textwidth]{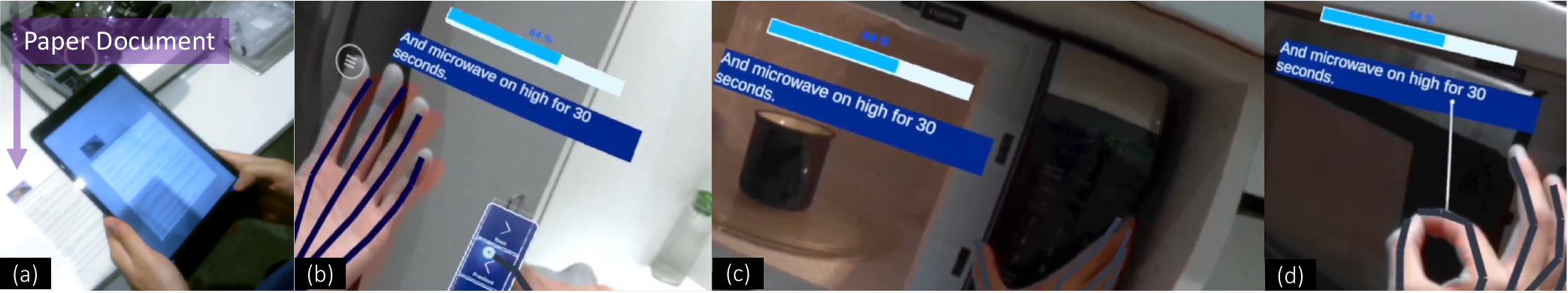}
    \vspace{-0.25in}
    \caption{Overview of \sysname: (a)~The author creates an MR experience by taking a snapshot of a paper document, with an optional ML-supported pipeline for associating key objects with each instruction step; (b)~The consumer can browse the spatialized instruction steps using a hand menu; (c)~The step is placed at an optimal location to minimize context switching and prevent occlusion of important interaction areas (\eg~not occluding the touchpad while setting the time on a microwave); (d)~The consumer can ``pinch-and-drag'' the step to refine the system placement. Steps (b - d) show the first-person MR view.}
    \Description{Figure 1.a show that the author could create an instruction MR experience by taking a snapshot of a paper document. The figure shows an demonstrative author attempt to taking a picture of an instruction document placed on the countertop in an office kitchen, using an iPad; Figure 1.b shows that the consumer could browse the spatialized instruction step using a hand menu. A demonstrative consumer is inspecting the instruction step ‘and microwave on high for 30 seconds’ after clicking the virtual hand menu attached to the left hand; Figure 1.c demonstrates that the instruction step would be ‘and microwave on high for 30 seconds’ is hovering in front of the user to minimize the context switching while preventing the occlusions of the importance interaction areas, such as the keyboard and display; Figure 1.d shows that the demonstrative consumer could “pinch-and-drag” the virtual instruction step if the system fails to find the optimal step placement.}
    \label{fig::teaser}
\end{teaserfigure}

\keywords{Mixed Reality (MR), Instruction Document, Tutorial and Help Systems, Context--Aware MR}

\author{Chen Chen}
\authornote{This work was started while Chen was interning at Adobe Research in summer 2022.}
\orcid{0000-0001-7179-0861}
\affiliation{%
  \institution{University of California San Diego}
  \city{La Jolla}
  \state{CA}
  \country{United States}
}
\email{chenchen@ucsd.edu}

\author{Cuong Nguyen}
\orcid{0000-0001-9234-9960}
\affiliation{%
  \institution{Adobe Research}
  \city{San Francisco}
  \state{CA}
  \country{United States}
}
\email{cunguyen@adobe.com}

\author{Jane Hoffswell}
\orcid{0000-0002-9871-4575}
\affiliation{%
  \institution{Adobe Research}
  \city{Seattle}
  \state{WA}
  \country{United States}
}
\email{jhoffs@adobe.com}

\author{Jennifer Healey}
\orcid{0000-0002-5700-4921}
\affiliation{%
  \institution{Adobe Research}
  \city{San Jose}
  \state{CA}
  \country{United States}
}
\email{jehealey@adobe.com}

\author{Trung Bui}
\orcid{0000-0002-0871-349X}
\affiliation{%
  \institution{Adobe Research}
  \city{San Jose}
  \state{CA}
  \country{United States}
}
\email{bui@adobe.com}

\author{Nadir Weibel}
\orcid{0000-0002-3457-4227}
\affiliation{%
  \institution{University of California San Diego}
  \city{La Jolla}
  \state{CA}
  \country{United States}
}
\email{weibel@ucsd.edu}

\renewcommand{\shortauthors}{Chen~\etal}

\begin{document}

\title[\sysname]{\sysname: Transforming Instruction Documents into Spatialized and Context-Aware Mixed Reality Experiences}

\begin{abstract}
While paper instructions are a mainstream medium for sharing knowledge, consuming such instructions and translating them into activities can be inefficient due to the lack of connectivity with the physical environment.
We propose \sysname, a novel workflow comprising an {\it authoring pipeline}, which allows the authors to rapidly transform and spatialize existing paper instructions into an MR experience, and a {\it consumption pipeline}, which computationally places each instruction step at an optimal location that is easy to read and does not occlude key interaction areas.
Our evaluation of the authoring pipeline with $12$ participants demonstrates the usability of our workflow and the effectiveness of using a machine learning based approach to help extract the spatial locations associated with each step.
A second within-subjects study with another $12$ participants demonstrates the merits of our consumption pipeline to reduce context-switching effort by delivering individual segmented instruction steps and offering hands-free affordances.
\end{abstract}

\maketitle

\section{Introduction}\label{sec::introduction}
Paper-based instructions are common for knowledge sharing.
Such instructions are often related to tasks that require users to interact with multiple objects spatially distributed in an environment.
For example, when following a recipe, a user may need to interact with multiple kitchen appliances like the cooktop, fridge, and microwave. %
When following a safety manual, a compliance manager may need to interact with various machines on the factory floor.

However, performing a task while consuming instructions can be tedious as the text is typically \textit{disassociated} from the user's physical environment. 
Thus, a user has to balance reading the instructions, figuring out what they mean in the environment, and performing the task, which can be cognitively demanding~\cite{Eiriksdottir2011}.
For example, when frying a piece of steak while following a cookbook, one needs to frequently switch between the cookbook and the pan to check the searing technique, temperature,~\etc 
This switch can be costly if the user places the cookbook somewhere peripheral so that it does not obstruct the task area. 
The user might end up spending more time trying to navigate the text and environment, than performing the task.
This problem is made worse if the user forgets some important information like temperature or duration, and has to repeatedly come back to the instruction to double check. 

Consumer Augmented Reality~(AR) and Mixed Reality~(MR)\footnote{While AR and MR are frequently interchangeable in literature, our contribution lies under the MR paradigm that focus on interactivity and context awareness~\cite{Skarbez2021}.} offer a unique opportunity to address this document-activity disassociation by overlaying digital elements on top of the environment.
These approaches are becoming more accessible, and studies have demonstrated their potentials for training workers to conduct tasks that are spatial in nature~\cite{Guntur2020}.
While prior works investigated the affordances of virtual guidance for conducting spatial tasks~\cite{Johnson2021}, and how to integrate such guidance in MR~\cite{Chidambaram2021}, they have not explored how document contents and their associated consumption experience could be constructed in MR. 
To that end, Microsoft Dynamic~$365$ Guides~\cite{Dynamic365} is an industry solution to help enterprise users manually create instructions and anchor them in an MR experience.

Across these MR instruction systems, the placement of the virtual instructions is often decided by the authors, and cannot be dynamically adapted to real-world contexts. 
This assumption could lead to undesirable experiences in both the consumption phase and the authoring phase. 
For the consumption phase, a static MR instruction could be mistakenly placed at a location too far from the user's task, at a distance that is difficult to read, or at a position that occludes key objects that the user is interacting with. 
For authoring, the author has to spend time associating and placing an instruction with its corresponding physical object. This process is time consuming and has to be repeated for every new set of instructions, even though the physical layout of objects might not change much over time.
While some prior works, \eg~FLARE~\cite{Gal2014Flare}, showed the usefulness of creating a persistent AR layout, real-world instructional activities are frequently changing (\eg~users might move from one place to another depending on the procedural step in the instructions), causing such a static layout to be infeasible.

We propose \sysname, a novel end-to-end workflow that transforms paper instructions into a context-aware instructional MR experience by {\it segmenting} monolithic documents; {\it associating} instruction steps with real-world anchoring objects; and optimally {\it placing} the virtual instruction steps so that they are easy to read and revisit while completing the tasks.
To realize this goal, \mbox{\sysname} consists of an {\it authoring} and a {\it consumption} pipeline.
With the \mbox{authoring} pipeline, the author can simply take a snapshot of the \mbox{paper} document by leveraging a mobile camera~(Fig.~\ref{fig::teaser}a).
Our system then segments the text in the document into individual instruction steps. 
The author can manually edit these segments, and \textit{associate} each step with the spatial location where the relevant activities will occur.
To help with this association task, we designed a machine learning~(ML) approach, where a fine-tuned language model was used to suggest the relevant spatial locations to the author. 
Our consumption pipeline is designed to render these instruction steps and place them in the associated spatial locations.
To ensure users could easily consume the spatialized instruction, the placement of each step in MR is optimized using a probabilistic optimization approach based on pre-created environmental models and the tracked gaze and hands.
Fig.~\ref{fig::teaser}c shows an example of a spatialized instruction step, where the step ``\textit{And microwave on high for 30 seconds}'' is tagged with ``{\it microwave}'' and is optimally rendered in front of the user while not occluding their view as they set the heating~time.

We prototyped \sysname~on Meta Quest Pro~\cite{QuestPro}, and conducted two within-subjects studies to evaluate the authoring pipeline with $12$ participants, and the consumption pipeline with another $12$ participants.
We demonstrated the usability of our authoring pipeline, and the effectiveness of using an ML-based approach to help the authors extract the spatial location associated with each step.
We then illustrated the effectiveness of the consumption pipeline for reducing context-switching effort, delivering the segmented instruction steps, and offering hands-free affordances.

With the assumption that the spatial profiles (\ie~the environmental geometry and associated semantic labels, see Sec.~\ref{sec::system_overview::design_insights}) are available, we contribute the design and evaluations of:

\vspace{4px}

\noindent$\bullet$ {\bf An authoring pipeline} that allows users to transform paper instructions into a spatialized MR experience;

\noindent$\bullet$ {\bf A consumption pipeline} that can computationally place the virtual instruction steps in the optimal place without either occluding the user's view or leading to large degrees of context switching.

\section{Related Work}\label{sec::related}
This paper is motivated by prior work on incorporating instruction experiences into MR and designing context-aware MR experiences.

\subsection{Integrating Instruction Experiences into MR}\label{sec::related::instruction_doc}
MR has been widely used for augmenting document consumption experiences~\cite{Qian2022, Chen2022VRContour, Chen2022VRContourWIP}.
Augmenting instructional documents, however, is still challenging due to the need to connect and integrate with real-world scenes and activities~\cite{Ganier2004, Schriver1997}.
Many prior works have explored the use of MR to augment a procedural instruction experience --- an important asynchronous collaboration task.
For example, ProcessAR~\cite{Chidambaram2021} proposed \insitu procedural AR instructions that could be rapidly created by experts, and used to teach novices through spatial and temporal demonstrations (\eg information about how to move a tool in the temporal domain and orient it in the spatial domain).
However, the placement of textual instructions was not explored.
CAPturAR~\cite{Wang2020} introduced a MR tool that helps users rapidly author context-aware applications, by referring to recorded activities.
Commercial tools such as Microsoft Dynamic~$365$ Guides~\cite{Dynamic365} enables experts to author a MR instruction experience by enacting the guidance, placing the instruction in the designated space, and recording the tool operations.

Although these works explored the design of MR-based instruction experiences, existing paper instructions are usually left behind, resulting in unnecessary time and effort to redesign a usable instruction workflow.
Additionally, existing MR instruction experiences are often not able to dynamically adapt to the changing environmental context of real-world activities (\eg~\cite{Dynamic365, Chidambaram2021}), causing user frustration when virtual graphics occlude interaction tasks.
\sysname~is novel in that it supports reusing existing paper documents that are designed by professional writers in a reader-centered way~\cite{Weller1986}, and can transform such documents into a spatialized and context-aware MR experience that is adapted to both the user's needs and the environmental characteristics.

\subsection{Designing Context-Aware MR Experiences}\label{sec::related::context_ar}
Context-aware MR systems aim to show {\it ``the `right' information, at the `right' time, in the `right' place, in the `right' way''}~\cite{Fischer2012}, which requires understanding both human and environmental contexts.

Prior research explored various computational approaches to realize this goal.
For example, Lindlbauer~\etal~\cite{Lindlbauer2019} used the real-time cognitive load, estimated by pupil dilation, to decide when and where the application should be shown, as well as how much information should be delivered (\ie~level of detail) in an MR system.
Lang~\etal~\cite{Lang2019} used simulated annealing to place virtual agents by considering key anchoring surfaces in the environment identified by pre-trained mask R-CNN.
Liang~\etal~\cite{Liang2021} used a similar approach to build a scene-aware virtual pet which could behave naturally in the real-world (\eg~respond to a food bowl).
Yu~\etal~\cite{Yu2021} proposed an interactive and context-aware furniture recommendation MR system by considering the real-world scene (\ie~spatial context) and the learnt furniture compatibility in a latent space (\ie~category context).
ScalAR enabled designers to author a semantically adaptive AR experience~\cite{Qian2022ScalAR}. 
Liu~\etal~\cite{Liu2021} attempt to generate suggestions for the arrangement of work surfaces in HoloLens, by capturing users' habitual behaviors of interacting with objects on the work surface.
Similar to our work, SemanticAdapt~\cite{Cheng2021} used an optimization approach to automatically adapt MR layouts between different environments by considering the virtual-physical semantic connections.
However, the target applications were only related to information consumption (\eg~consuming news feeds) and did not consider those related to real-world activities. 

Inspired by this prior work, \sysname~demonstrates a novel consumption pipeline that leverages a similar computational approach to analyze the tracked gaze and hand position, as well as the anchoring surfaces of the key objects in a target environment.

\section{Preliminary Needs-Finding Study}\label{sec::prelim}
To understand the pain-points for consuming existing paper instructions (\ie~{\it monolithic}, \textit{non}-segmented, \textit{non}-spatialized, and \textit{non}-context-aware), we conducted a needs-finding study, rooted in participant observations~\cite{Bogdewic1992, Spradley2016} and semi-structured interviews.

\subsection{Participants, Tasks, and Procedure}\label{sec::prelim::procedures}
We recruited four participants (age, $M = 24.75$, $SD = 2.87$, two~females, two~males).
Participants were required to complete the designated task using paper instructions.
Specifically, PP1 and PP4 were required to make coffee with a coffee machine by following the user manual~\cite{CoffeeManual}~(Fig.~\ref{fig::prelim_study_tasks}a).
PP2 and PP3 were required to make a chocolate microwave cup cake from an existing recipe~\cite{ChocolateMugCake}~(Fig.~\ref{fig::prelim_study_tasks}b).
These tasks were chosen since they are common activities in an office kitchen; require participants to read the textual instructions; and could be completed in a reasonable time for an unpaid study.
We also used the existing instructions~\cite{CoffeeManual, ChocolateMugCake} created by professional writers to minimize the impact of non-professional writing styles.
All participants reported little (PP3, PP4) to no (PP1, PP2) prior knowledge for the designated task.
Next, we conducted a semi-structured interview, focusing on {\it ``what are the pain-points while performing the designated tasks using the given paper instruction, and why?''}~
Finally, we brainstormed potential designs of a MR experience for consuming instruction documents. 
We explained the concept of MR for PP3 and PP4 who were not familiar with it.
Participants were encouraged to sketch their imagined design on an iPad Canvas.
The study took on average $30$~min ($SD = 3.25$~min).

\subsection{Findings}\label{sec::prelim::findings}
Overall, we identified three pain-points through the study.

\begin{figure}[t]
    \centering
    \includegraphics[width=0.485\textwidth]{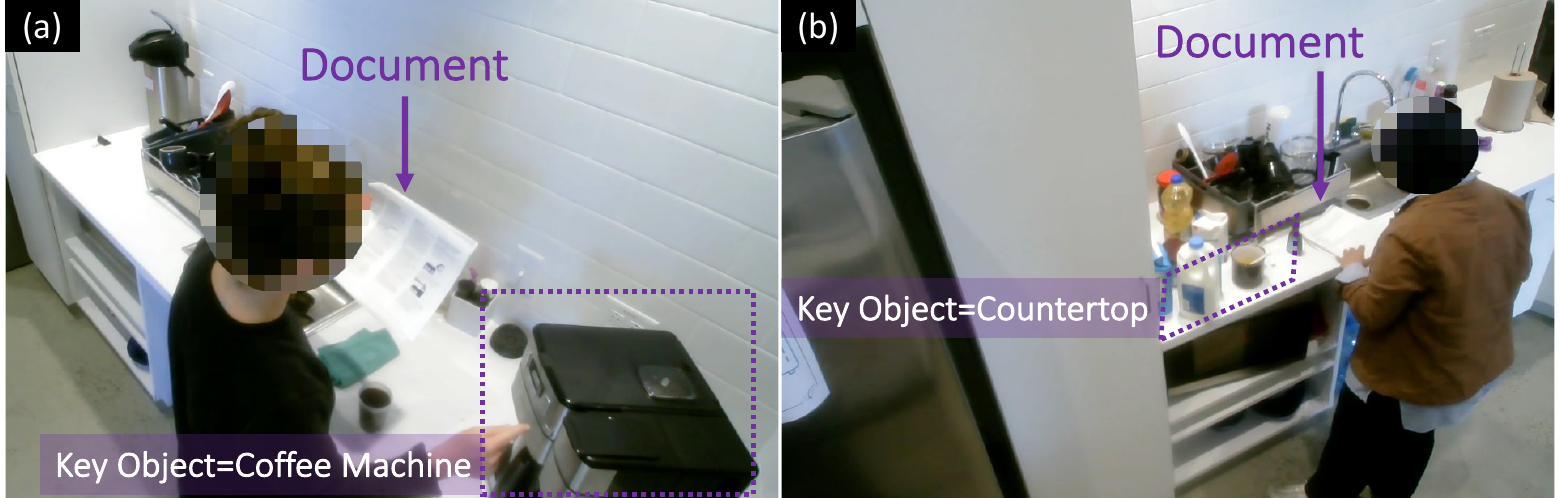}
    \vspace{-0.3in}
    \caption{Preliminary needs-finding tasks. (a)~Making a cup of coffee using a coffee machine (PP1). (b)~Making a chocolate cake in a mug with a microwave (PP3).}
    \Description{Figure 2.a shows the preliminary needs-finding tasks of PP1, who was making a cup of coffee using a coffee machine in an office kitchen. Figure 2.b demonstrates the preliminary needs-finding tasks of PP3, who was making a chocolate cake in a mug with a microwave (PP3) in an office kitchen.}
    \label{fig::prelim_study_tasks}
    \vspace{-0.20in}
\end{figure}

\begin{figure}[t]
    \centering
    \includegraphics[width=0.485\textwidth]{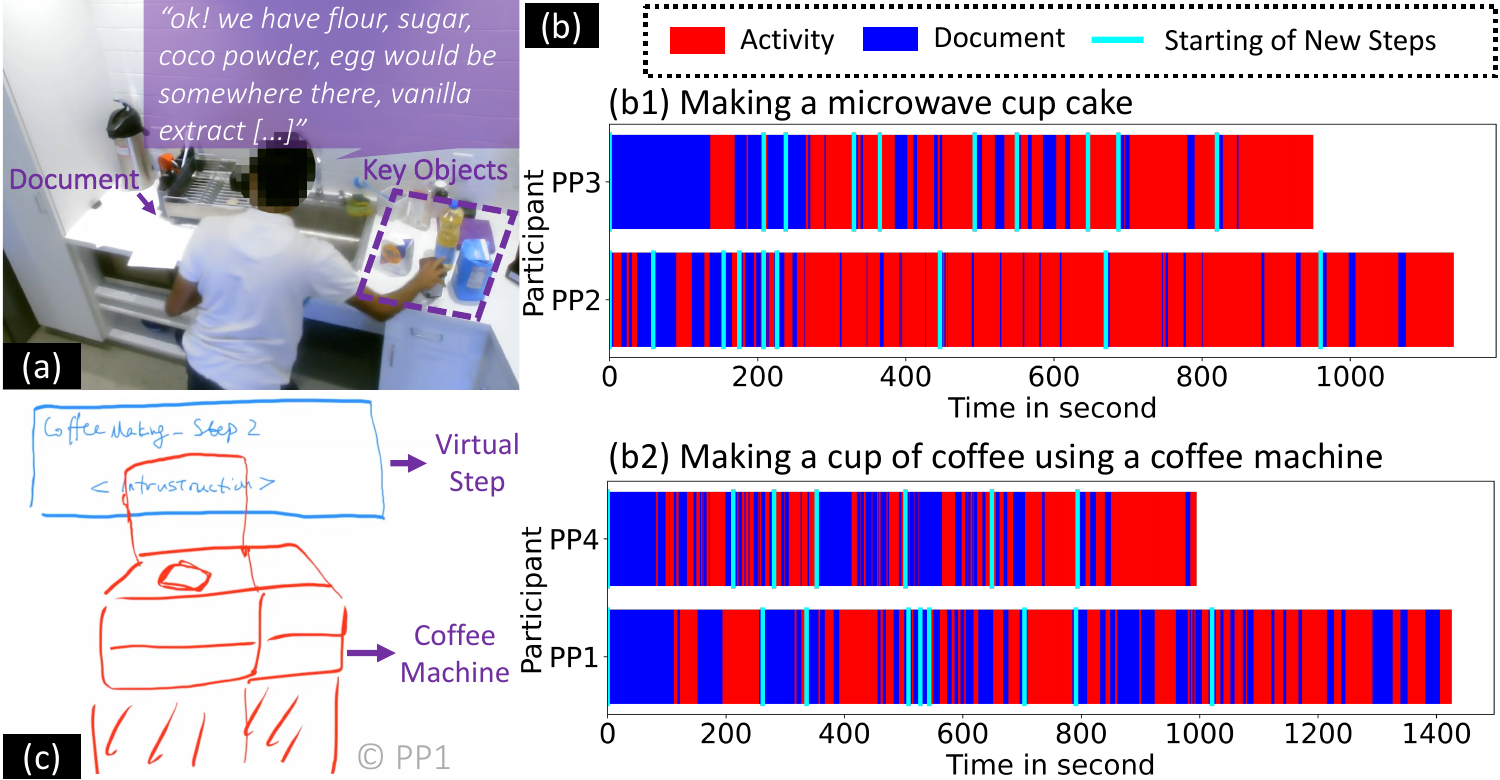}
    \vspace{-0.25in}
    \caption{Preliminary needs-finding study results. (a) Example context switching while PP2 was attempting to map instructions with real-world objects. (b) The annotated timestamps showing participants' current focus as either the document or real-world activities. (c) PP1's design for an instructional MR experience while using a coffee machine.}
    \Description{Figure 3.a shows the example context switching while PP2 was attempting to mentally ``map'' the instructions with real-world objects. While completing the tasks, PP2 was also commenting: ``ok! we have flour, sugar, cocoa powder, egg would be somewhere there, vanilla extract [...]''; Figure 3.b1 demonstrates the annotated timestamps showing participants’ current focus as either the document or real-world activities. This subplot shows the annotated results in time domain while PP2 and PP3 were making a microwave mug cake; Figure 3.b2 describes the annotated timestamps showing participants’ current focus as either the document or real-world activities. This subplot shows the annotated results in time domain while PP1 and PP4 were making a cup of coffee using a coffee machine. Figure 3.c describes the PP1’s design for an instructional MR experience while using a coffee machine. A virtual step is attached at the top of the physical coffee machine.}
    \vspace{-0.2in}
    \label{fig::prelim_study_results}
\end{figure}

\vspace{4px}
\noindent{\bf The overwhelming amount of information or lack of necessary details in the instructions can impact the usability.}
During the semi-structured interviews, two participants~(PP1, PP3) pointed out that the sometimes overwhelming amount of information and irrelevant content could be distracting. 
For example: {\it ``there was a lot of information on the document. And it wasn't easy for me to know where and what information I should be looking at''}~(PP1) and {\it ``the first setback was too much information''}~(PP3).
To handle the potentially overwhelming amount of information, PP2 first skimmed the document in search of relevant content: 
{\it ``I am first attracted to see where the bullet points are. [...] And then if I just skim through the first two or three points, I understand that this is not relevant. So I'm just skipping those sections completely.''} 
While designing the MR experience, PP1 incorporated such insights into her design (Fig.~\ref{fig::prelim_study_results}c) and commented:
{\it ``I would rather the MR just gives me one small step every time. For example, I am making coffee and reaching the step two, and in the virtual instruction, it will say like coffee making step two, and then here will be just a instruction with just a few sentences.''}

On the other hand, PP2 and PP4 believed the lack of details for certain steps could impact their ability to perform the sub-task.
For example: {\it ``I've never cracked an egg so I don't know how to do it. [...] If it is some things that I've never actually done [and the instruction document does not tell me how], I might actually be confused''}~(PP2).
In essence, different users likely require different levels of information based on their prior experience with the intended task.

\vspace{4px}\noindent{\bf There are missing connections between the instruction step and real-world activities.}
All participants identified a need for establishing spatial connections between instructions and real-world objects or activities.
For example, during the design phase of the study, PP2 emphasized that {\it ``having [a virtual] arrow [in MR] that can help me and connect me to the object is helpful.''}

On the other hand, most participants suggested that overwhelming spatial guidance might be unnecessary and could lead to visual disturbance, similar to their concerns about the potentially overwhelming amount of information already in some instructional documents.
Reflecting further on PP2's suggestion to include virtual arrows, he highlighted some potential downsides of this approach, noting that
{\it ``[having useless spatial indicators] is going to be overloaded. I know these basic things, and I don't need pointers to see like a spoon or a mug.''}~
PP1's design sketches implied an alternative approach to establish connections between the instruction step and real-world objects by placing the virtual step close to the coffee machine, without occluding the user's view (Fig.~\ref{fig::prelim_study_results}c).

\begin{figure*}[t]
    \centering
    \includegraphics[width=\textwidth]{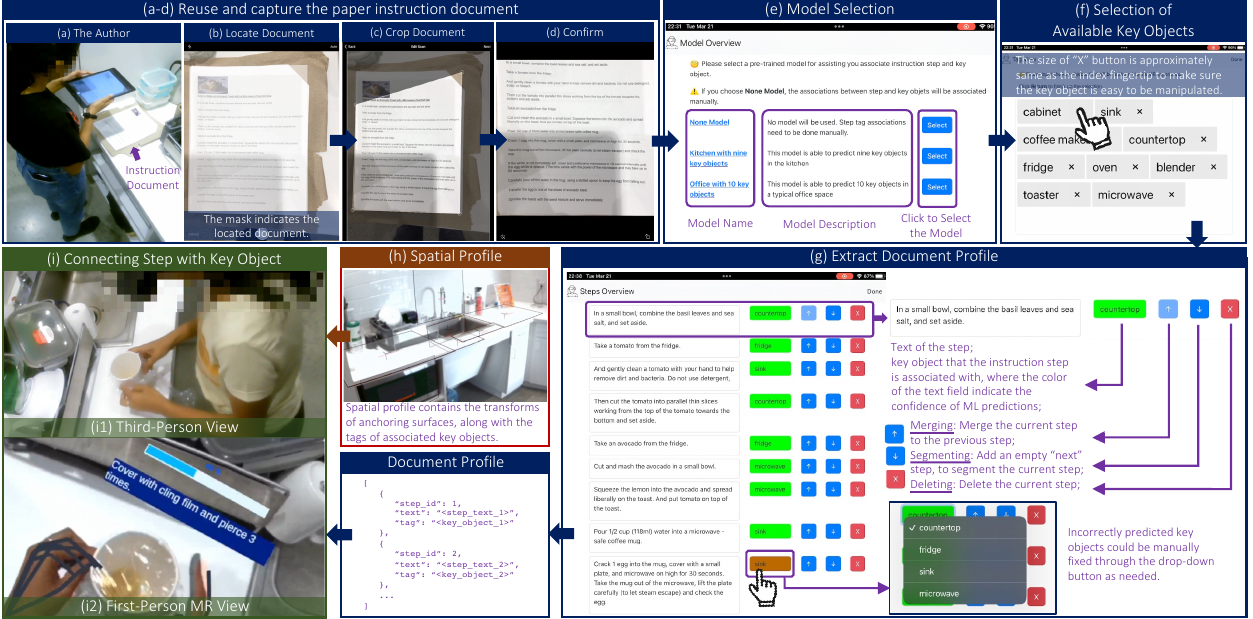}
    \vspace{-0.25in}
    \caption{\sysname~system overview. We assume a spatial profile \mbox{(h, \red{red} block)} was pre-created. The author uses the authoring pipeline \mbox{(a - g, \blue{blue} blocks)} to extract the document profile for the MR experience. With consuming pipeline \mbox{(i, \green{green} block)}, the instruction steps are displayed based on the environmental (loaded via the spatial profile) and user's contexts.}
    \Description{Figure 4.a shows that an author participant was holding an iPad and attempting to take a picture of a paper instruction on the countertop in an office kitchen; Figure 4.b describes the screenshot from the iPad with a semi-transparent mask, which identified the location of the paper document; Figure 4.c shows the screenshot of the iPad showing that the semi-transparent mask could be cropped by the authors; Figure 4.d describes the screenshot of the iPad showing the final cropped document; Figure 4.e shows the screenshot of the iPad that enables the authors to select the model from three available models: ``None Model'', ``Kitchen with nine key object'', and ``Office with 10 key objects''; Figure 4.f describes the screenshot of the iPad with a set of tags, including: ``cabinet'', ``sink'', ``coffee machine'', ``countertop'', ``fridge'', ``oven'', ``blender'', ``toaster'', and ``microwave''; Figure 4.g shows the screenshot of the iPad that allows the author to re-segment the instruction steps, or extract the key object that is associated with each instruction step. Key components in this interface have been highlighted; Figure 4.h describes the kitchen environment with the key anchoring surfaces being highlighted by semi-transparent masks. In this example, a semi-transparent surface is attached to the microwave; two semi-transparent surface is attached to the countertop; and four semi-transparent surface is attached to the sink; Figure 4.i1 illustrates the third-person view of a participant wearing a Quest Pro headset attempting to cover the cup with cling film; Figure 4.i2 shows the first-person view of the Figure 4.i1, where an instruction step ``cover with cling film and pierce 3 times'' is hovering right above the cup.}
    \vspace{-0.10in}
    \label{fig::system_overview}
\end{figure*}

\vspace{4px}
\noindent{\bf Frequent context-switching between the instruction document and real-world activities should be minimized.}
By analyzing the video recording, we found that users perform frequent context-switching between documents and real-world activities.
For example, while PP2 was conducting the task, he naturally commented: {\it ``ok! we have flour, sugar, coco powder, egg would be somewhere there, vanilla extract [...]''}, thereby demonstrating frequent context-switching between instructions and real-world objects  as he checked off items from the recipe list~(Fig.~\ref{fig::prelim_study_results}a).
Participants generally used two types of strategies to minimize context-switching: {\it holding} and {\it switching to} the document while performing the task.
Specifically, we found that PP1 tended to hold the documents while performing tasks, although occasionally this approached was inconvenient for steps requiring two hands (Fig.~\ref{fig::prelim_study_tasks}a).
In contrast, other participants tended to place the document on the countertop while reading the content, and move the document to a new place when switching to other steps (Fig.~\ref{fig::prelim_study_tasks}b).
To understand how participants used the documents to perform individual steps, we manually labelled the timestamp of the recorded video while participants were switching between documents and real-world activities, or otherwise. 
Through this process, we observed that the participants highly relied on the documents to perform tasks. 
Specifically, while participants were attempting to complete a particular step, the document was still frequently referenced even though the participants had already read it at the beginning of each step (Fig.~\ref{fig::prelim_study_results}b).

\subsection{Design Considerations}\label{sec::prelim::considerations}
We identified three fundamental \underline{D}esign \underline{C}onsiderations (DCs) by analyzing the data from our preliminary needs-finding process.

\vspace{4px}\noindent
{\bf [DC1] Only delivering the segmented instruction could enhancing information consumption experience.}
We show that the participants expect to consume relevant information corresponding to their current activities, yet existing paper documents usually deliver all information to users at the same time.
An improved MR instruction consumption experience could create novel and flexible ways to segment the document, such that only relevant information is delivered to users for each associated step.

\vspace{4px}\noindent
{\bf [DC2] Optimally placing instruction texts next to the areas of interactive activities might be helpful for the intermittent and repetitive information consumption experience.}
While participants generally read the entirety of an instruction step before performing the corresponding actions, the instruction step may need to be consumed {\it repeatably} (Fig.~\ref{fig::prelim_study_results}b).
Therefore, the placement of the virtual instruction step in MR should consider the spatial location where the relevant task would occur.

\vspace{4px}\noindent
{\bf [DC3] The right level of spatial guidance could help users associate instructions with spatialized key objects.}
While few existing works~(\eg~\cite{Johnson2018, Johnson2021}) suggested the usefulness of spatial guidance for MR-based instructional experiences, our participants emphasized the importance of {\it moderate} spatial guidance, with limited visual disruptions.
Thus, usable spatial guidance, without causing overwhelming visual disturbance, should be provided at the beginning of each instruction step.

\section{\sysname~System Overview}\label{sec::system_overview}
Based on \S\ref{sec::prelim::findings}, we designed \sysname~(Fig. \ref{fig::system_overview}), comprising of two pipelines:
\textbf{(1)}~{\bf an authoring pipeline} for an {\bf author} to rapidly and easily create a spatialized MR instruction experience from existing paper-based instructions (Fig.~\ref{fig::system_overview}a - g,~\S~\ref{sec::authoring}), 
and \textbf{(2)}~{\bf a consumption pipeline} for enabling a {\bf consumer} to explore context-aware, spatialized instruction steps in MR (Fig.~\ref{fig::system_overview}i,~\S~\ref{sec::consuming}).
While we use cooking tasks in an office kitchen as a running example for the design and evaluations, our approach could be transferred to other types of instruction documents.

\subsection{Assumptions and System Walkthrough}\label{sec::system_overview::design_insights}
We consider an {\bf environment} to be a typical workspace (\eg~the kitchen) for supporting a procedural {\bf task}~(\eg~baking a cake). 
Each environment contains multiple physical {\bf key objects}, which are defined as the important, stationary objects that are usually attached to the environment permanently (\eg~the fridge and microwave).
We did not consider non-stationary objects (\eg~a~mug) due to the lack of support for real-time arbitrary object tracking with Quest Pro~\cite{QuestPro}.
Each key object contains one or multiple {\bf anchoring surfaces}, which are virtual surface(s) that describe the approximated geometry of the objects.
We use these surfaces to determine the placement of instruction in MR.

\begin{figure}[b]
    \centering
    \vspace{-0.2in}
    \includegraphics[width=0.485\textwidth]{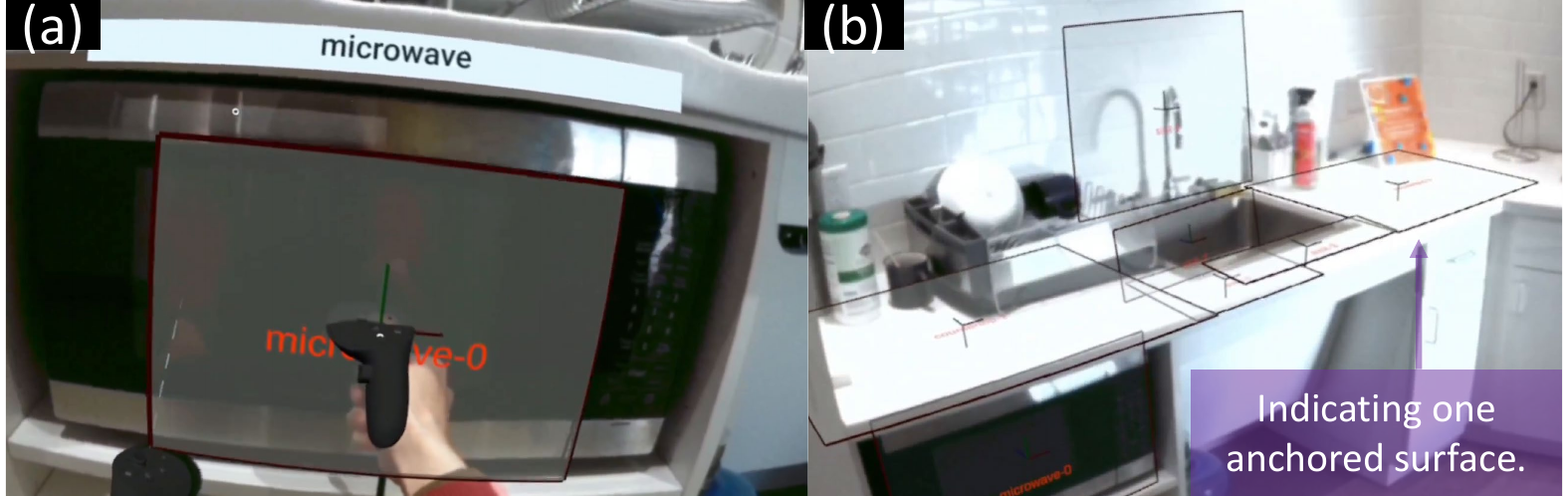}
    \vspace{-0.25in}
    \caption{(a)~Creation of an anchoring surface, visualized as a semi-transparent mask, using touch controllers. (b)~Examples of anchoring surfaces in our experimental kitchen. Both scenes were captured as first-person MR views.}
    \Description{Figure 5a shows the first-person view of creation of an anchoring surface, visualized as a semi-transparent mask, using touch controllers; Figure 5b describes the examples of anchoring surfaces in our experimental kitchen. Both scenes were captured as first-person MR views. In this example, a semi-transparent surface is attached to the microwave; two semi-transparent surface is attached to the countertop; and four semi-transparent surface is attached to the sink.}
    \label{fig::anchoring_surfaces}
\end{figure}

We also define a \textbf{spatial profile} as a collection of the labels of these key objects and their anchoring surface(s). 
We assume that a spatial profile could be created offline, either through automatic geometry processing (\eg~\cite{AppleRoomPlan}) or manual means (Fig.~\ref{fig::anchoring_surfaces}).
Example environment where spatial profiles could be created in advance include office kitchen, a rental house, and a factory.
To realize this assumption, we implemented a MR interface in the Quest Pro~\cite{QuestPro} that allow each anchoring surface (represented by a 2D plane) and the associated key object could be easily declared (Fig.~\ref{fig::anchoring_surfaces}a). 
The spatial anchor APIs~\cite{OculusSpatialAnchors} were used to ensure the declared anchoring surfaces are persistent in the environment~(Fig.~\ref{fig::anchoring_surfaces}b).

With these assumptions, the goals of the two pipelines of \sysname~are described as below:

\vspace{4px}\noindent$\bullet$~{\bf Authoring Pipeline.}
Given an existing paper-based instruction document, the author would first segment the document content into smaller steps, each of which will only be associated with one key object.
For each step, the author needs to identify the {\bf metadata}, including: \textbf{(i)} the text of each instruction step, and \textbf{(ii)} the key object that the step should be anchored on.
Together, such metadata makes up our \textbf{document profile} --- the essential elements to recreate a spatialized MR experience from existing paper documents.

\vspace{4px}\noindent$\bullet$~{\bf Consumption Pipeline.}
While consuming the document in MR, the instruction steps float in mid-air and will be optimally attached to one of the anchoring surfaces of the associated key object, based on {\bf user context}---real-time user interactions data in MR such as the tracked eye gaze and hand joints.
For example, the step {\it ``boil a cup of water in the microwave for 5 min''} should be attached to one of the anchoring surfaces of the {\it``microwave''}~key object and not impact the user's interactions.

\subsection{Application Scenarios}\label{sec::system_overview::users}
We consider two user roles, where the {\bf author} use the authoring pipeline to rapidly create an instruction MR experience and the {\bf consumer} use the consumption pipeline to consume the authored MR experience while completing tasks. 
We target on colocated and asynchronous collaborations~\cite{Johansen1988}, where the instructions are authored and consumed in same environment.
Specific application scenarios are described below.

\vspace{4px}
\noindent{\bf Author and Consumer are Different Users.}
For a specific procedural task, \sysname~could be used to facilitate asynchronous collaborations between experts and novices.
For example, company administrators could use \sysname~for training new employees to use the provided facilities (\eg~coffee machines and fridges in the shared office kitchen). 
Chidambaram~\etal~\cite{Chidambaram2021} also demonstrated the usefulness of using a similar instruction MR experience to teach novices assembly mechanics.

\vspace{4px}\noindent{\bf Author and Consumer are the Same User.}
While we differentiate two user roles, it is possible that the author and consumer are the {\it same} user.
As the cognitive processes for consuming instructions usually occur in working memory, which is constrained by both time and processing capacity, it is often necessary for one to revisit the procedures repeatably for the {\it same} task~\cite{Ganier2004, Anderson2013}. 
For example, while cooking the same meal, it is common for the user to refer back to the cookbook each time upon starting a new step. 
Fig.~\ref{fig::prelim_study_results}b confirmed such patterns, where participants repeatedly refer to the instructions while completing a specific step.
In this scenario, the user could author a personalized and spatialized MR experience, which s/he could use repeatably when preparing the same meal in the future; for example, a user could customize a recipe to take into account his/her preferences for spice level.

\section{Authoring Pipeline}\label{sec::authoring}

\noindent The authoring pipeline extracts the document profile from an existing paper document to create an MR experience {\it rapidly} and~{\it easily}.

\subsection{Document Capture and Parsing}\label{sec::authoring::snapshot}
\noindent One question for document reuse is {\it how to enable users to rapidly capture and extract the document profile from an existing instruction document}?
Inspired by mobile applications that allow users to capture and analyze scanned documents (\eg~Adobe Scan~\cite{AdobeScan} and Tab~\cite{Tab}), we similarly enable authors to simply take a snapshot of the instruction document to generate a document~profile (Fig.~\ref{fig::system_overview}a - d).

The author can then adjust the scanned region to crop out unnecessary components (\eg~titles, \etc) as needed (Fig.~\ref{fig::system_overview}c). 
\sysname~then leverages OCR services by Google Vision API~\cite{GoogleVisionAPIOCR} to parse the scanned image into machine readable text, due to its ability to extract paragraph structure in the parsed text using full text annotations~\cite{GoogleVisionAPIFullTextAnnotation}.
By default, we segment each paragraph as one step in the instructions. However, the author can re-segment the steps and fix errors in a dedicated mobile~interface (Fig.~\ref{fig::system_overview}g).

\subsection{Selecting the Model and Key Objects}\label{sec::authoring::selecting_key_object}
To extract the document profile, we designed a manual and ML-assisted approach to help authors rapidly and easily associate key objects with each step.
Our ML-assisted approach leverages a pre-trained language model for a specific environment to predict the key object that is associated with each step.
After transforming the existing paper document into machine readable text, the authors need to select the model for the target environment (Fig.~\ref{fig::system_overview}e).
For environments without a pre-trained model, the manual approach enables authors to manually extract the metadata of each step.

The author then selects the key objects that exist in the target workspace (\ie~the set of {\it available} key objects) (Fig.~\ref{fig::system_overview}f). 
First, this step ensures the key objects contained in the extracted document profile aligns with the spatial profile of the intended environment. 
For example, a cooking instruction step such as {\it ``boil a cup of water''} could be executed either in a typical household kitchen on a {\it cooktop}, or in an office kitchen that only provides a {\it microwave}. 
Second, setting the available key objects also provides prior knowledge to help increase the accuracy while predicting key object associations.
For example, if an office kitchen only has a {\it microwave}, the key object for {\it ``boil a cup of water''} should not be predicted as {\it oven}, even though {\it oven} is a possible label for the pre-trained model (\S~\ref{sec::authoring::edit}).

\subsection{Creating Document Profile}\label{sec::authoring::edit}
Creating a document profile requires two types of metadata:

\vspace{4px}\noindent{\bf (1) The text of each procedural step.}
While by default we consider and segment each sentence as one instruction step (\S\ref{sec::authoring::snapshot}), the author could overwrite the system segmented results by {\it segmenting}, {\it merging} and {\it deleting} specific step(s) (Fig.~\ref{fig::system_overview}h). 
When a specific step is modified or a new merged step is generated, the associated key objects will be re-predicted (if the ML supported mode is used).
Although some instruction steps might be associated with multiple or no key objects, such flexibility allows the author to split or merge the target step(s).
In response to \textbf{[DC1]}, additional flexibility is also provided for the author to modify the generated text of each step to ensure that the right information with right level of details could be delivered to the consumers.

\vspace{4px}\noindent{\bf (2) The key object that the step is associated with.}
The authors can use either a manual or ML-assisted approach to determine the key object associated with each step.
Fig.~\ref{fig::system_overview}g shows a dedicated interface with segmented instructions, where the authors can use the drop downs to select the associated key objects, with the color scale indicating the confidence of our ML predictions (if applicable).

While manually assigning each step to a key object could work robustly, predicting key objects using a ML-assisted approach that requires a pre-trained model is challenging, due to the needs for a dataset and ground truth labels.
Creating such a dataset that could be generalized to {\it all} procedural instructions is not realistic, and labeling each step with a ground truth key object is also difficult and time consuming.
Instead of preparing such dataset for \textit{all} instruction documents, we chose to focus on domain-specific dataset that is publicly available.
Alternatively, the dataset could be created via vendors or crowdsourcing.

We describe our methods for generating such a pre-trained model below.
Although our running example is based on cooking instructions, the overall approach could be transferred to other type of instructions documents, provided the unlabeled dataset is available. 

\begin{figure}[t]
    \centering
    \includegraphics[width=0.5\textwidth]{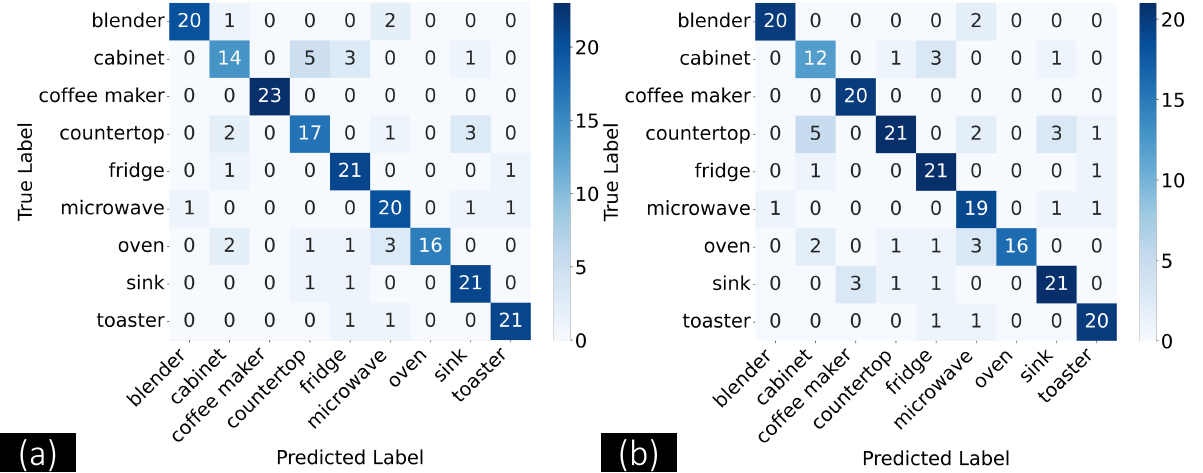}
    \vspace{-0.25in}
    \caption{Confusion matrices of the fine-tuned BERT model, with the ground truth generated by rule-based method (a) and manually labeling (b).}
    \Description{Figure 6.a describes the confusion matrices of the fine–tuned BERT model, with the ground truth generated by rule-based method. The numbers from top to bottom include: ``blender, 20, 1, 0, 0, 0, 2, 0, 0, 0'', ``cabinet, 0, 14, 0, 5, 3, 0, 0, 1, 0'', ``coffee maker, 0, 0, 23, 0, 0, 0, 0, 0, 0'', ``countertop, 0, 2, 0, 17, 0, 1, 0, 3, 0''; ``fridge, 0, 1, 0, 0, 21, 0, 0, 0, 1'', ``microwave, 1, 0, 0, 0, 0, 20, 0, 1, 1''; ``oven, 0, 2, 0, 1, 1, 3, 16, 0, 0''; ``sink, 0, 0, 0, 1, 1, 0, 0, 21, 0''; ``toaster, 0, 0, 0, 0, 1, 1, 0, 0, 21''; Figure 6.b demonstrates the confusion matrices of the fine–tuned BERT model, with the ground truth generated by manually labeling. The numbers from top to bottom include: ``blender, 20, 0, 0, 0, 2, 0, 0, 0''; ``cabinet, 0, 12, 0, 1, 3, 0, 0, 1, 0''; ``coffee maker, 0, 0, 20, 0, 0, 0, 0, 0, 0''; ``countertop, 0, 5, 0, 21, 0, 2, 0, 3, 1''; ``fridge, 0, 1, 0, 0, 21, 0, 0, 0, 1''; ``microwave, 1, 0, 0, 0, 0, 19, 0, 1, 1''; ``oven, 0, 2, 0, 1, 1, 3, 16, 0, 0'', ``sink, 0, 0, 3, 1, 1, 0, 0, 21, 0''; ``toaster, 0, 0, 0, 0, 1, 1, 0, 0, 20''.}
    \vspace{-0.10in}
    \label{fig::model_performance}
\end{figure}

\vspace{4px}\noindent{\bf Dataset:}
We used RecipeNLG~\cite{Bien2020RecipeNLG} for training purposes, which contains more than $2.2$M cooking recipes where each recipe includes multiple ordered instruction steps.
The process of aggregating all steps from all recipes yielded an unlabelled dataset with $19.5$M steps, with an average of $11.54$ ($SD = 7.13$) words~per~step.

\vspace{4px}\noindent{\bf Rule-Based Labelling of Training Instruction Steps:} 
Instead of manually labelling each step, we used a rule-based approach to label steps that contain the \textit{exact} words of the predefined key objects.
For example, we label the step {\it ``boiling a cup of water in the \underline{microwave} for 5 min''} as {\it ``microwave''}, yet the step {\it ``boiling a cup of water for 5 min''} will not be labeled, and thus will not be included as part of our training dataset.
We iteratively selected nine key objects that exist in a typical kitchen: {\it blender}, {\it cabinet}, {\it coffee maker}, {\it countertop}, {\it fridge}, {\it microwave}, {\it oven}, {\it sink}, {\it toaster}.
Such selections also ensures a reasonable amount of instruction steps for subsequent model fine-tuning purposes.
We generated a dataset where each of the nine labels is associated with $218$ instruction steps (\ie~the \emph{labeled} dataset contains $218 \times 9 = 1962$ instruction steps).

\vspace{4px}\noindent{\bf Training:}
Due to the limited dataset size, we fine-tuned the model using the output classification layer of a $12$-layer BERT model for uncased vocabulary, which has been used for generating contextual language embeddings~\cite{Devlin2018}.
We used Adam optimizer with the learning rate, $\epsilon$, and batch size set to $2\times10^{-5}$, $10^{-8}$, and $32$, respectively, recommended by Devlin~\etal~\cite{Devlin2018}.
$80\%$, $10\%$ and $10\%$ of the dataset are used for training, validation and testing, with the amount of steps for each key object balanced across the sets.

\vspace{4px}\noindent{\bf Model Performance:}
We demonstrated an overall $83.57\%$ testing accuracy by considering the label generated by our rule-based approach as the ground truth (Fig.~\ref{fig::model_performance}a).
Additionally, we manually label the associated key objects on the testing dataset to limit the impact from any errors generated by our rule-based approach, which lead to a $82.13\%$ overall accuracy (Fig.~\ref{fig::model_performance}b).

\vspace{4px}\noindent{\bf Model Execution: }
The pre-trained model is used for predicting key objects from the segmented text of each step. 
To enhance the accuracy of the predicted key objects, we use prior knowledge provided while specifying the available key objects (\S\ref{sec::authoring::selecting_key_object}).
Specifically, the final assigned key object is the ML predicted key object with the highest confidence score that also belongs to the set of available key objects of the target environment.

\setlength{\belowdisplayskip}{0pt} \setlength{\belowdisplayshortskip}{0pt}
\setlength{\abovedisplayskip}{0pt} \setlength{\abovedisplayshortskip}{0pt}
\setlength{\textfloatsep}{10pt}

\section{Consumption Pipeline}\label{sec::consuming}
The consumption pipeline aims to spatialize each steps by anchoring them at the optimal position next to the key object.
For example, consider how the instruction {\it``microwave on high for 30 seconds''} should be attached to a microwave. An ideal location would be at the front surface of the microwave door. A less idea location would be at the front of the input panel because the instruction might get in the way when the user tries to set the timer (see examples in Fig.~\ref{fig::teaser}c, Fig.~\ref{fig::system_overview}i, and Fig.~\ref{fig::consumption::examples} in Appendix~\ref{sec::app::study_results::consuming}).

\subsection{Interaction Design}\label{sec::consuming::interactions}
Our consumption pipeline provides dedicated interaction metaphors based on the preliminary findings (\S\ref{sec::prelim::findings}).

\vspace{4px}\noindent{\bf Navigating Between Individual Steps.}
Consumers can use hand menus to easily and rapidly switch between steps (Fig.~\ref{fig::teaser}b).
We adopted the suggestions from \textbf{[DC1]} and the conceptual design of Fig.~\ref{fig::prelim_study_results}c that advocate the idea of delivering the right level of information only at the right time.
Therefore, \sysname~only renders the current instruction step along with a task completion progress bar.
When a new step is triggered, \sysname~first anchors the virtual label in front of the consumer, since a initial instruction step consuming is usually required before consumers proceeding on execute the associated steps (Fig.~\ref{fig::prelim_study_results}b).

\vspace{4px}\noindent{\bf Animating Spatial Guidance.}
To address \textbf{[DC3]}, we decided not using the persistent visual guidance (\eg~virtual arrows)~\cite{Johnson2018, Johnson2021} that might cause unnecessary visual disturbance.
Instead, we use a animated flying effect where the virtual step could {``fly''} toward the key object after initial instruction step consuming.
Such design leverage the fact that a motion effect could direct the consumers' attention, and could implicitly and rapidly offer visual guidance of the spatialized key object without causing overwhelming disturbance while consumers are executing the steps~\cite{Harley2014}.

\vspace{4px}\noindent{\bf Placement of Instruction Steps.}
\sysname~places and anchors the instruction step on one of the anchoring surfaces of the key objects while not occluding the important region.
This design emphasizes \textbf{[DC2]} suggesting the connections between instruction and real-world contexts, and could bring convenience while the consumers are attempting to refer back to the instruction step repeatedly while completing the step (Fig.~\ref{fig::prelim_study_results}c).
If the consumer dislikes the label position, they can request a new position update on-demand using a mid-air pinch gesture.
We also allow the consumers to use pinch-and-drag gestures to manually move the step to their preferred place (Fig.~\ref{fig::teaser}d).
Such feedback action will in turn help on future decisions while placing instruction steps.

\begin{table}[t]
    \small
    \centering
    \begin{tabular}{ll} 
     \hline
     \textbf{Notations} & \textbf{Descriptions} \\
     \hline\hline 

     $W$, $H$ & \makecell[l]{Number of discretized cells along the width and \\ height of the anchoring surface.} \\ 
    
     $rot_s$ & \makecell[l]{Rotation, represented by quaternion, of the surface $s$.}  \\
     
     \makecell[l]{$du_s$, $dr_s$, $df_s$} & \makecell[l]{The up, right, and forward direction of the surface $s$.}\\
     
     $p_{eye}$ & \makecell[l]{The midpoint of left and right eye in world coordinate.}\\ 
     
     $df_{eye}$ & \makecell[l]{The forward direction of the gaze, averaged by left \\ and right eye gaze.}\\ 
      
     $a = (r, c, s)$ & \makecell[l]{Representation of an instruction step placement with \\ respect to anchoring surface $s$, with index of $r$ and $c$ \\ along width and height, where $r \in [0, W)$, $c \in [0, H)$.}\\
     
     $p_{a}$ & \makecell[l]{The position of the instruction step in world coordinate.}\\

     $\theta_{x, y}$ & \makecell[l]{The angle between vector $x$ and $y$.}\\
     
     \hline
    \end{tabular}
    \caption{Notations of the key parameters and functions.}
    \label{table:notations}
\end{table}

\subsection{Problem Formulation}\label{sec::consuming::problem}
We formulated the process of optimally placing the instruction step as an optimization problem, where we used the tracked hands and gaze, as well as the anchoring surfaces defined by the spatial profile to search the optimal placement for each step.
Table~\ref{table:notations} summarizes the notations of key parameters.

\vspace{4px}\noindent{\bf Representations of an Instruction Step Placement.}
We first discretized each anchoring surface into $W \times H$ virtual cells where each cell has a dimension of $3cm \times 3cm$.
We assumed that the center of each virtual step should be aligned with the center of the cell on the anchoring surface.
We used $a = (r, c, s)$ to indicate the placement of a step, where $r$ and $c$ indicate the index of cell along the width and height of the surface $s$.
The world position of the attempted step placement is $p_{a} = p_{topLeft} + 0.03 \cdot dr_s \cdot r - 0.03 \cdot du_s \cdot c$, where $p_{topLeft}$ is the position of the top left vertex of surface $s$. 

\begin{figure}[t]
    \centering
    \includegraphics[width=0.485\textwidth]{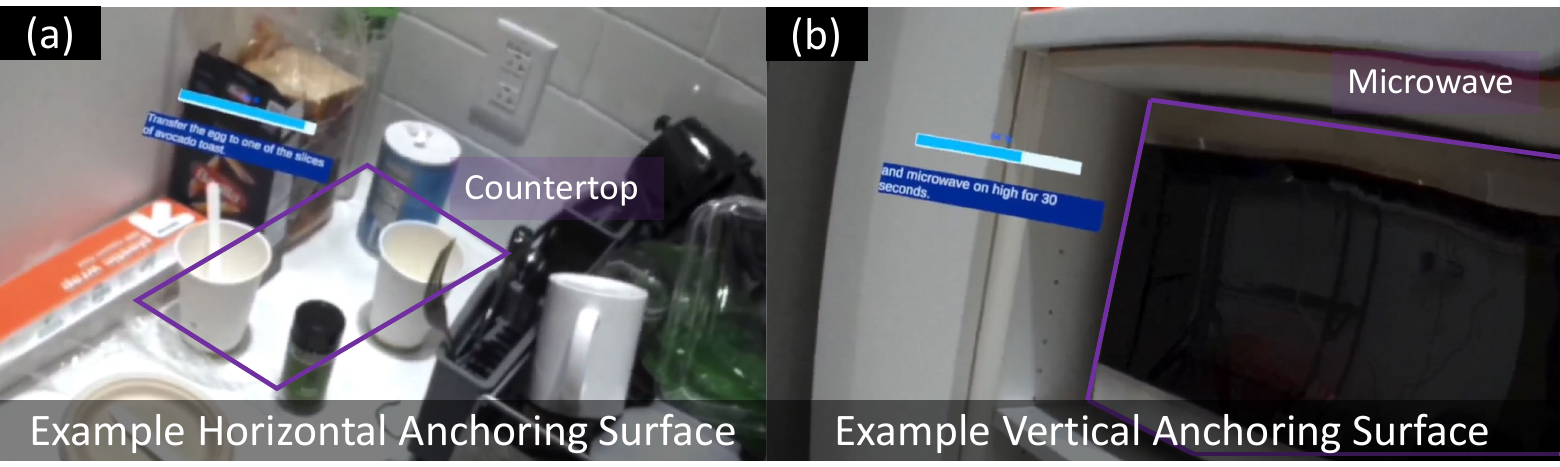}
    \vspace{-0.25in}
    \caption{First-person view of rotating instruction step for (a) horizontal anchoring surface (\eg~countertop) and (b) vertical anchoring surface (\eg~microwave).}
    \Description{Figure 7.a describes the first-person view of rotating instruction step for horizontal anchoring surface (e.g., countertop). Figure 7.b shows the first-person view of rotating instruction step for vertical anchoring surface (e.g., microwave).}
    \vspace{-0.1in}
    \label{fig::label_rotation}
\end{figure}

\vspace{4px}\noindent{\bf Representations of the Step Label Rotation.}
Reading angles is a critical factor for consuming document~\cite{Morris2007}.
It is therefore important to determine the rotation of the step, such that the text is always perpendicular to user's looking direction (\ie~the virtual texts should be delivered facing toward user's eye).
To address this, we rotated the anchoring surfaces by using the potential looking direction~($p_{a} - p_{eye}$).
Algo.~\ref{alg:lable_rot} shows how we compute the rotation of the step for horizontally (\eg~countertop) and vertically placed anchoring surfaces (\eg~the front surface of fridge).
Fig.~\ref{fig::label_rotation} demonstrates two examples where the steps are appropriately rotated.

\begin{algorithm}[t]
   \small 
   \caption{Computing the rotation of instruction step.} \label{alg:lable_rot}
   \begin{algorithmic}[1]
        \PROCEDURE{GetRotation}{$rot_{s}$, $p_a$, $p_{eye}$}
            \STATE $dir \gets p_a - p_{eye}$  \COMMENT{Approximate potential looking direction.}
            \STATE $\alpha_{up} \gets Angle(du_s, dir)$
            \STATE $\alpha_{right} \gets Angle(dr_s, dir)$           
            \IF{$s$ is a horizontal anchoring surface}
               \STATE \textbf{return} $rot_{s} * Quaternion.Euler(90^{\circ} - \alpha_{up}, 0, 90^{\circ} - \alpha_{right})$
            \ELSE[$s$ is a vertical anchoring surface.]
                \STATE \textbf{return} $rot_{s} * Quaternion.Euler(90^{\circ} - \alpha_{up}, \alpha_{right} - 90^{\circ}, 0)$
            \ENDIF
        \ENDPROCEDURE
    \end{algorithmic}
\end{algorithm}

\vspace{4px}\noindent{\bf Importance Map.}
One goal while attempting to place an instruction step onto associated anchoring surface(s) is to find the optimal placement that will not occlude consumer's interactions with the key object.
While Lang~\etal~\cite{Lang2019} presumed that the centroid of a key object is the most important area and should not be occluded by virtual MR agents, such assumption is invalid in our problem as the real-world interactions are highly dynamic.
For example, while interacting with microwave, the critical areas might be the region on top of keypad when inputting cooking time, or the center areas when the user is checking whether the food is cooked.
We used {\bf importance map}~($imap$) to represent the importance of each possible cell on anchoring surface, where a larger $imap$ value implies a higher probability that the corresponding position being interacted, and therefore should not be occluded by the virtual step. 
The values of $imap$ are determined using near real-time data provided by the MR headset.
By leveraging the pre-created spatial profile, we used the tracked gaze and hands to infer the importance of each possible position on anchoring surface(s).
Intuitively, the areas near the hands, which usually imply the regions that are interacted by the consumers, might be more important and should not be occluded by the steps.
\S\ref{sec::consuming::map_and_surface} describes the computations of $imap$.

\subsection{Importance Map on Anchoring Surfaces}\label{sec::consuming::map_and_surface}

\begin{algorithm}[t]
   \small 
    \caption{Approximating the importance of individual frame.}
    \label{alg:frame_weights}
    \begin{algorithmic}[1]
        \PROCEDURE{GetFrameWeights}{$v_{left}[\;]$, $v_{right}[\;]$, $t[\;]$}
            \STATE $N \gets Length(t)$  \COMMENT{$N$ indicates the frame size.}
            \STATE $w, w_{speed}, w_{time} \gets [0] * N, [0] * N, [0] * N$
        
            \FOR{$i \gets 1$ to $N$}
                \STATE $w_{speed}[i] \gets (|v_{left}[i]| + |v_{right}[i]|)/2$
                \STATE $w_{time}[i] \gets (t[i]- t[1]) / (t[N] - t[1]) $
            \ENDFOR
        
            \STATE $w_{speed} \gets MinMaxNormalize(w_{speed}) $
            \COMMENT{Normalize to $0$ to $1$.}
            
            \STATE $w_{time} \gets MinMaxNormalize(w_{time}) $
            \COMMENT{Normalize to $0$ to $1$.}
            
            \STATE $w \gets Normalize((1 - w_{speed}) w_{time}) $ \COMMENT{$\Sigma_1^{N}(w[i])$ should be $1$.}
        
            \STATE \textbf{return} {$w$}
        
        \ENDPROCEDURE
    \end{algorithmic}
\end{algorithm}

\begin{figure*}[t]
    \centering
    \includegraphics[width=\textwidth]{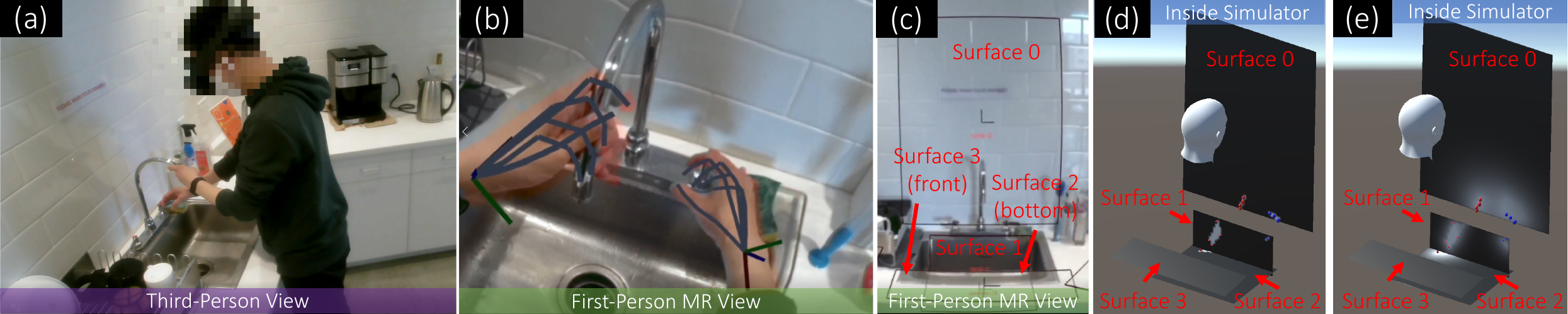}
    \vspace{-0.25in}
    \caption{Examples of importance map. (a) Third-person view; (b) First-person view through MR; (c) First-person view of the key object (sink) containing four anchoring surfaces; (c) Importance map from single frame; (d) Overall importance map from a set of frames. The \red{red} and \blue{blue} points in (d) and (e) indicate the projected key joints of the tracked left and right hand on the anchoring surfaces, respectively.}
    \Description{Figure 8.a shows a third-person view of a participant using the sink in an office kitchen; Figure 8.b describes the first-person view through MR where the participant’ hand joints are tracked; Figure 8.c shows the sink with four anchoring surfaces, each of which is highlighted by a semi-transparent mask; Figure 8.d describes the importance map from a single frame, captured by the Meta Quest Pro. The areas that are occluded by the hands on surface 0, 1, and 2 are visualized by white area; Figure 8.e shows the overall importance map from a set of frames.}
    \vspace{-0.15in}
    \label{fig::importance_map}
\end{figure*}

\noindent{\bf Approximate the Importance of Each Frame.} 
The contextual data from each {\bf frame} refers to the tracked gaze and hand joints, at a specific time instant.
Inferring $imap$ from contextual data on single frame is less reliable, as real-world activities are highly dynamic.
Yet, contextual data collected from a set of frames are practically not equally important.
Therefore, while aggregating contextual data across a set of frames, it is important to consider the relative importance of individual frames.
\sysname~uses the tracked eye behaviors to approximate the relative importance of contextual data of each frame.
The \first intuition is based on the instantaneous angular speed of gaze, where a slower angular speed of gaze implies a higher importance.
For example, while attempting to input time when using microwave, the consumer might be \textit{fixing} at the keypad, during which the contextual data could offer meaningful clues for approximating $imap$.
Whereas, the consumer might rapidly \textit{saccade} around the environment while finding ingredients, during which the contextual data might be less meaningful.
While \cite{Nystrom2010, Mould2018} attempted to design closed-form solution for classifying saccade and fixation using the speed of gaze, such eye behaviours usually varies across users and tasks.
Our \second intuition is based on the observation that the contextual data from a more recent frame might be more useful to indicate the interactions in the subsequent task episode.
Algo.~\ref{alg:frame_weights} shows the computation of the {\bf frame weight}~($w$) that is used to quantify the relative importance of contextual data at each frame.
Experimentally, we set $N = 90$, which is approximately $1$ second of past contextual data.
We approximated $w$ based on instantaneous angular speed of left and right gaze~($v_{left}, v_{right}$), and the timestamps ($t$) of each frames.
Remarkably, a slower eye moves (\ie~smaller $v_{left}$ and $v_{right}$) and a more recent timestamp (\ie~ larger $t$) would lead to a more important frame weight.
The $MinMaxNormalize(\cdot)$ and $Normalize(\cdot)$ represent the {\it min}-{\it max} normalization (to $[0, 1]$), and the normalization process such that the summation of the list is $1$.

\begin{algorithm}[t]
    \small
    \caption{Algorithms for computing importance map.}
    \label{alg:map}
    
    \begin{algorithmic}[1]

        \PROCEDURE{GetMap}{$trackedJoints[]$, $W$, $H$}
            \STATE $hits, mask \gets [], zeros(W, H)$
            \FOR{$p_{joint}$ in $trackedJoints$}
                \STATE $hit \gets RayCast(start:p_{eye}, direction:p_{joint} - p_{eye})$
                \IF{$hit \neq null$}
                    \STATE $w \gets ToWidthIndex(hit.x)$
                    \STATE $h \gets ToHeightIndex(hit.y)$
                    \STATE $mask[w, c] \gets 1$
                    \STATE $hits.add([w, c])$
                \ENDIF
            \ENDFOR
            
            \STATE $p_{hull} \gets ConvexHull(hits)$
            \STATE $mask \gets FloodFill(mask, p_{hull})$
            \STATE \textbf{return} {$mask$}
        \ENDPROCEDURE
        \STATE  
        \PROCEDURE{GetOverallMap}{$frames$, $W$, $H$}
        
            \STATE $map \gets zeros(W, H)$
            \FOR{$i \gets 1$ to $N$}
                \STATE $imap_{left} \gets GetMap(frames[i].left.joints, W, H)$
                \STATE $imap_{right} \gets GetMap(frames[i].right.joints, W, H)$
                \STATE $map \gets map + w[i] * (Soft(imap_{left}) + Soft(imap_{right}))$
            \ENDFOR
            \STATE \textbf{return} {$MinMaxNormalize(map)$}     \COMMENT{Normalize to $0$ to $1$.}
        \ENDPROCEDURE
    \end{algorithmic}
\end{algorithm}

\vspace{4px}\noindent{\bf Approximate the Importance Map from Tracked Hands.}
To compute the overall $imap$ from a set of frames, we first approximated the $imap$ from the contextual data collected at each frame.
This could be realized by $GetMap(\cdot)$ (Algo.~\ref{alg:map}), which $FloodFill$ the $ConvexHull$ of the projected points of $15$~tracked hand joints~\cite{OculusHandTracking} on the associated anchoring surfaces. 
Fig.~\ref{fig::importance_map}d shows an example of the $imap$ from single frame, approximated by tracked hands in Fig.~\ref{fig::importance_map}b.

\vspace{4px}\noindent{\bf Generate the Overall Importance Map.} 
While we assumed that the areas that are not occluded by hands might be less important, the importance assigned to each cell on the anchoring surfaces might not be equally same.
For example, while inputting time using keyboard of microwave, the further the step being placed from the areas occluded by hands could lead to lower probability that the important areas being occluded by hands. 
Therefore, $imap$ should be expected to model {\it how important of a specific pixel on the anchoring surface}, instead of {\it whether the particular pixel is important}.
Inspired by Lang~\etal~\cite{Lang2019}, we used $Soft(\cdot)$ to approximate the importance at each possible placement on the anchoring surface(s) generated by left and right hand respectively (Algo.~\ref{alg:map}). 
Eqn.~\ref{alg:blur_map} describes this process, where $e_{min}^{r, c}$ indicates the $L2$-distance from placement $(r, c)$ to the closest placement(s) where the computed importance of individual frame from $GetMap(\cdot)$ (Algo.~\ref{alg:map}) is $1$.

\begin{equation}
    \label{alg:blur_map}
    imap(r, c) = 1 - \frac{e_{min}^{r, c}}{max_{r', c'}e_{min}^{r', c'}}
\end{equation}

The overall $imap$ is finally computed by aggregating the $imap$ generated by left ($imap_{left}$) and right ($imap_{right}$) hands on each frame using previous computed weight ($w$) in Algo.~\ref{alg:frame_weights}.
This process could be demonstrated in $GetOverallMap(\cdot)$ (Algo.~\ref{alg:map}).
Notably, if there is no area being occluded by hands, we set $imap = \bm{O}_{W, H}$. 
Fig.~\ref{fig::importance_map}e shows an example overall $imap$.

\subsection{Constraints and Costs}\label{sec::consuming::cost}
\noindent To solve the optimization problem for placing the instruction step on the anchoring surface(s), we need to model the constraints such that {\it the placed step will minimally occlude the user's view and will not be too far from the user's focused attention}.

\vspace{4px}\noindent{\bf Total Cost.} 
We designed the overall cost ($C_{total}(a)$, Eqn.~\ref{eq::total_cost}) as a weighted sum of the visibility cost ($C_V({a})$), the readability cost ($C_R({a})$), hand angle cost ($C_{HA}({a})$), and preference cost ($C_P({a})$).
$\lambda_V$, $\lambda_R$, $\lambda_{HA}$, and $\lambda_P$ are the weights associated with each of designed cost.
Experimentally, we set them to $0.24$, $0.24$, $0.24$ and $0.28$, with slight emphasis on consumers' preference.
We provide rationales of the design of each costs.

\begin{figure}[t]
    \centering
    \includegraphics[width=0.480\textwidth]{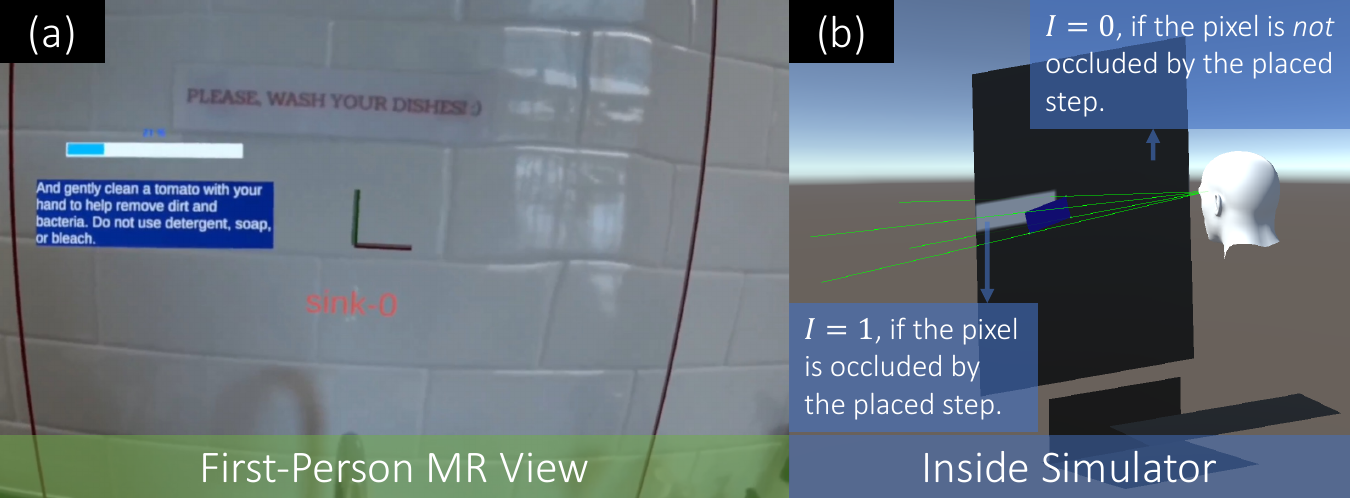}
    \vspace{-0.25in}
    \caption{Example of occlusion map. (a)~An example placement of the instruction step anchored next to the sink; (b)~Visualization of the generated occlusion map, see Fig.~\ref{fig::importance_map}c for the corresponding real-world scene.}
    \Description{Figure 9.a shows an example placement of the instruction step anchored next to the sink; Figure 9.b describes the visualization of the generated occlusion map, where the area being occluded is visualized by white area.}
    \vspace{-0.05in}
    \label{fig::occlusion_map}
\end{figure}

\begin{equation}
    C_{total}(a) = \lambda_V C_{V}(a) + \lambda_R C_{R}(a) + \lambda_{HA} C_{HA}(a) + \lambda_P C_{P}(a) 
    \label{eq::total_cost}
\end{equation}

\vspace{4px}\noindent{\bf (i) Visibility Cost.}
$C_V$ aims to to measure how much key areas of the anchoring surfaces are occluded by a step placement $a$, and to penalize the situation while the step occluding the important areas.
Eqn.~\ref{eq::visibility_cost} defines $C_V$, where $I$ indicates {\bf occlusion map} and $imap$ indicates the relative importance of each discretized cells on the anchoring surfaces computed by Algo.~\ref{alg:map}.

\begin{equation}
    C_{V}(a) = \frac{[\sum_{r, c} I(r, c) \cdot imap(r, c)]^2}{||imap(r, c)||_2 \cdot \sum_{r', c'}I(r', c')}  \label{eq::visibility_cost}
\end{equation}

Notably, the $I(r, c)$ is assigned as $1$ when the pixel $(r, c)$ is occluded by the step from the center of both eye (Fig.~\ref{fig::occlusion_map}).
Fig.~\ref{fig::occlusion_map}b shows an example occlusion map when the instruction step is placed next to the wall behind the sink.
We finally normalized $C_V(a)$ to be independent of dimensions of anchoring surface(s).

\vspace{4px}\noindent{\bf (ii) Readability Cost.}
We penalized the solution when the delivered step is too far from the user's attention.
To model this constraints, we used the eye tracking results and measure the $L2$-distance from the placed step to the weighted average of looking direction of both eye ($df_{eye}$).
Eqn.~\ref{eq::readability_cost} defines $C_R(a)$, where $d_{max}$ represents the maximum $L2$-distance between two arbitrary solutions on the anchoring surfaces, which is computed by the maximum distance of the convex hull consisting of all vertices of the anchoring surfaces.
Additionally, $C_R(a)$ need to enforce the instruction step is placed within binocular vision (\ie~approximately $\pm 60^{\circ}$)~\cite{Pettigrew1986}, to minimize the needs of moving head in order to read the instructions.
We used a coefficient $k$ to penalize the cost function, where $k$ is set to $1$ if $\theta_{p_a - p_{eye}, df_{eye}} < 60^{\circ}$, otherwise we set  $k = \theta_{p_a - p_{eye}, df_{eye}}$.

\begin{equation}
    C_{R}(a) = \frac{k \cdot ||({p}_{a} - {p}_{eye}) \cdot {df}_{eye} \cdot {df}_{eye} - ({p}_{a} - p_{eye})||_2}{d_{max}} \label{eq::readability_cost}
\end{equation}

\vspace{4px}\noindent{\bf (iii) Hand Angle Cost.}
We modeled the observation that the instruction documents are usually held by and placed in front of the consumers' hands (Fig.~\ref{fig::prelim_study_tasks}).
We first computed the angle between forward direction of the hand and the direction vector pointing from hand to the attempted solution ($\theta_{df_{hand}, p_{a} - p_{hand}}$), for left and right hand respectively (noted as $\theta_{leftHand}(i)$, $\theta_{rightHand}(i)$ computed from frame $i$).
We then formulated the overall hand angle cost, by aggregating the angle cost generated by each frame (Eqn.~\ref{eq::hand_angle_cost}).

\begin{equation}
    C_{HA}(a) = \sum_{i} \frac{w[i] \cdot [\theta_{leftHand}(i) + \theta_{rightHand}(i)]}{360^{\circ}} \label{eq::hand_angle_cost}
\end{equation}

\vspace{4px}\noindent{\bf (iv) Preference Cost.}
While \sysname~could determine the optimal step placement based on near-real-time context data, we retained the flexibility for users to specify their preferred placements.
Eqn.~\ref{eq::preference_cost} defines $C_P(a)$, where $p_{pref, S_i}$ refer to the preferred step placement in the world coordinates for step $S_i$.
We used $p_{pref, S_i} = null$ to indicate that the user has not manually fix the step placement.
Similar to $C_R$, we used $d_{max}$ to normalize preference cost.

\begin{equation}
    \label{eq::preference_cost}
     C_{P}(a) = 
    \begin{cases}
       \frac{||p_{a} - p_{pref, S_i}||_2}{d_{max}} & p_{pref, S_i} \neq null\\
       0 & p_{pref, S_i} = null
    \end{cases}
\end{equation}

\begin{figure}[t]
    \centering
    \includegraphics[width=0.47\textwidth]{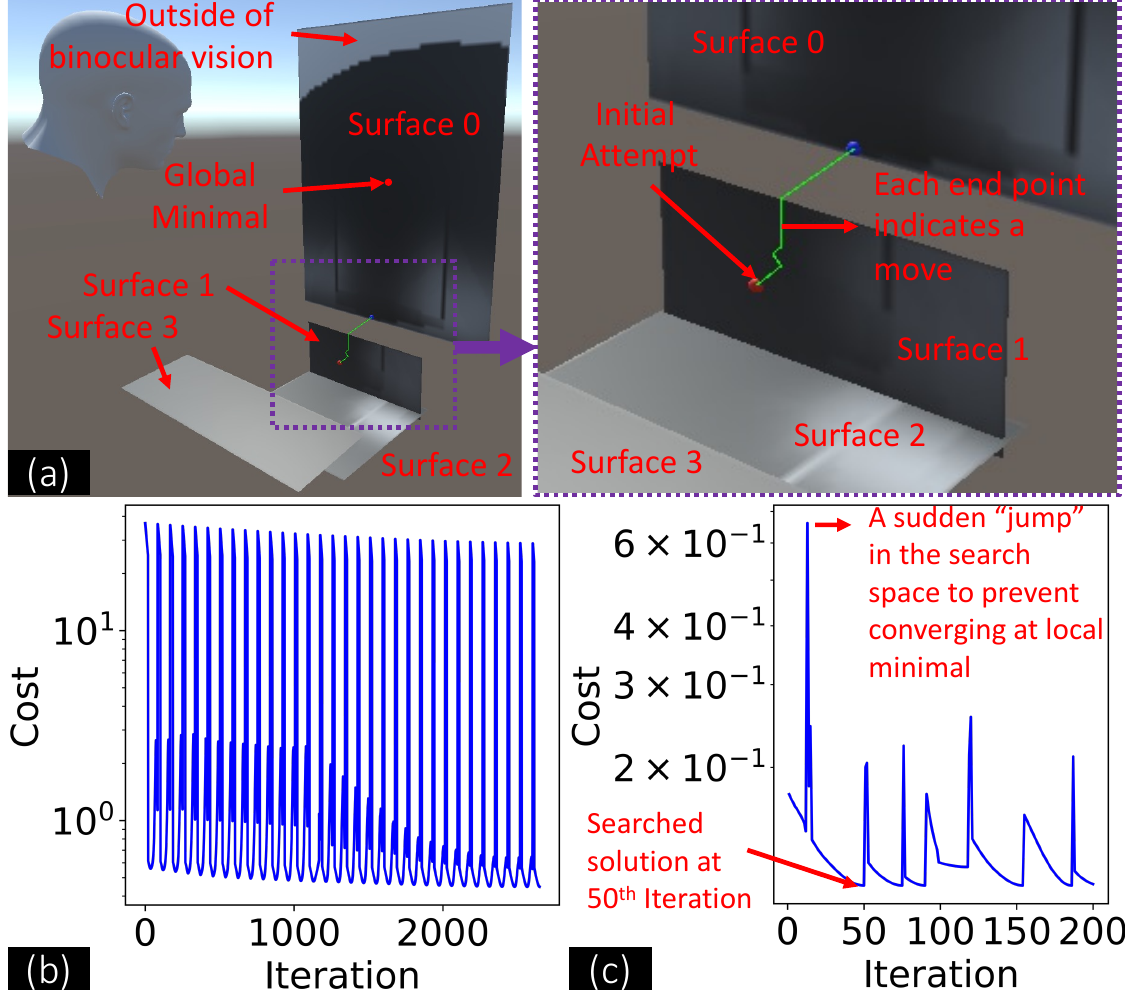}
    \vspace{-0.1in}
    \caption{(a) Example placement attempts (the green trace). The darker color of the anchoring surface indicates a lower $C_{total}$; (b - d) Example optimized cost over each iteration using greedy algorithm~(b) and simulated annealing approach~(c). To increase readability, $log$ scale is applied for $y$-axis. We only showed the traces before current cost reaching global minimal.}
    \Description{Figure 10.a shows the example placement attempts (the green trace). The darker color of the anchoring surface indicates a lower C_{total}. The placement corresponds to global minimum is approximately at the center of the surface 0; Figure 10.b describes an example optimized cost over each iteration using greedy algorithm. To increase readability, log scale is applied for y-axis.We only showed the traces before current cost reaching global minimal; Figure 10.c shows the example optimized cost over each iteration using simulated annealing approach. A random “jump” in the search space is highlighted by an annotation arrow, which aims to prevent converging at local minimum. To increase readability, log scale is applied for y-axis.We only showed the traces before current cost reaching global minimal.}
    \vspace{-0.05in}
    \label{fig::example_energy_traces}
\end{figure}

\begin{figure*}[t]
    \centering
    \includegraphics[width=\textwidth]{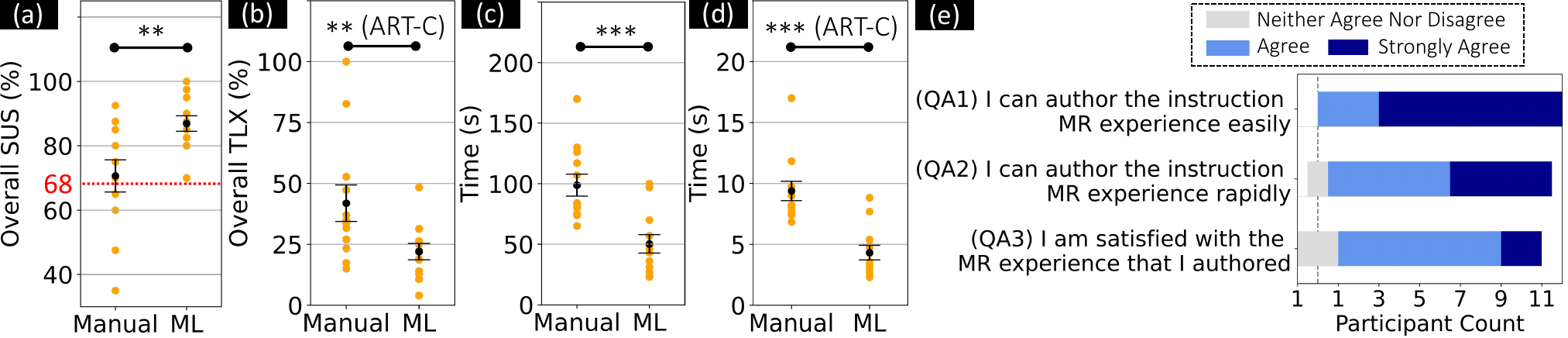} 
    \vspace{-0.3in}
    \caption{Results of authoring pipeline evaluations. (a - d) The overall SUS, weighted TLX scores, TCT of extracting document profiles, and the average task completion time for deciding each instruction step with segmented instruction step; (d) Survey results of how participants assessed the overall authoring pipeline and authored MR experience.}
    \Description{Figure 11.a shows the overall SUS scores of manual and ML based approach. ``**'' is annotated as the statistical significance; Figure 11.b shows the overall NASA TLX scores of manual and ML based approach. ``** (ART-C)'' is annotated as the statistical significance; Figure 11.c describes the task completion time of extracting document profiles of manual and ML based approach. ``***'' is annotated as the statistical significance; Figure 11.d shows the average task completion time of extracting document profiles for each instruction step of manual and ML based approach. ``*** (ART-C)'' is annotated as the statistical significance; Figure 11.d describes the survey results of how participants assessed the overall authoring pipeline and authored MR experience. For ``QA1 - I can author the instruction MR experience easily'', three participants and nine participants self-reported as ``agree'' and ``strongly agree''; For ``QA2 - I can author the instruction MR experience rapidly'', one, six, and five participants self-reported as ``neither agree nor disagree'', ``agree'', and ``strongly agree''; For ``QA3 - I am satisfied with the MR experience that I authored'', two, eight, and two participants self-reported as ``neither agree nor disagree'', ``agree'', and ``strongly agree'' with the statement.}
    \vspace{-0.15in}
    \label{fig::authoring_eval}
\end{figure*}

\subsection{Optimizations}
We aimed to minimize $C_{total}$ by searching optimal placement for a specific step $\hat{a}$ (\ie~$\hat{a} = \argmin_{a} C_{total}(a)$).
Finding optimal placement by computing $C_{total}$ for each possibilities is impractical due to unnecessary latency and computational overhead.
We instead used simulated annealing to approximate the global optimal~\cite{Kirkpatrick1983}.

\vspace{4px}\noindent{\bf Make a Move by Choosing the Neighbour Solution.}
We chose the neighbour solution $a^{i+1} = (r^i + {\delta_r}^{i}, c^i + {\delta_c}^i)$ by making a move of current solution $a^{i} = (r^i, c^i)$ at iteration $i$.
Eqn.~\ref{eq::move} describes our approach to determine the change over width (${\delta_r}^{i}$) and height (${\delta_c}^{i}$) of the anchoring surface at iteration $i$ that yields smallest $C_{total}$, where $({\delta_r}^{i}, {\delta_c}^{i}) \in \{({\delta_r}, {\delta_c}) | {\delta_r}, {\delta_c} \in \{ \pm1, 0\} \land {\delta_r}{\delta_c} \neq 0\}$.

\begin{equation}
    \delta_r, \delta_c = {argmin}_{\delta_r', \delta_r'} {C_{total} (a^{i + 1} = (r^i + \delta_r', c^i + \delta_r'))} 
    \label{eq::move}
\end{equation}

Notably, to ensure the sampled neighbouring solution is on the target anchoring surface, Eqn.~\ref{eq::move_constraints} should be satisfied.

\begin{equation}
    \label{eq::move_constraints}
        0 \leq r^{i+1} \leq W \quad 0 \leq c^{i+1} \leq H \quad r^{i+1}, c^{i+1} \in \mathbb{N}
\end{equation}

Although such method performs well while making moves \textit{within} single anchoring surface, the attempted solution will not be made \textit{across} the surfaces.
To address this, we specify that the placement on the neighbour anchoring surface, which is closest to the current placement attempt, would be chosen upon Eqn.~\ref{eq::move_constraints} being violated.  
Fig.~\ref{fig::example_energy_traces}a demonstrates an example of how the neighbour solution is chosen while placement moving across the surfaces.
To prevent converging at local minimal, we choose a random move in the global search space if $C_{total} (a^{i + 1}) > C_{total} (a^{i}) $~\cite{Kirkpatrick1983}.

\vspace{4px}\noindent{\bf Metropolis-Hastings Sampling and Simulated Annealing.}
We first selected a random placement on one of anchoring surface(s) randomly and used the Metropolis-Hastings algorithm~\cite{Robert1999} to sample the subsequent attempt.
The probability for accepting the new attempt $a^{i+1}$ is determined by Metropolis criteria.
Eqn.~\ref{eq::samping} defines the the transition kernel of the Markov chain, where $T(i)$ indicates the temperature that will decay over the iteration.

\begin{equation}
    p(a^{i+1} | a^i; i) = min \{1, exp(\frac{C_{total}(a^i) - C_{total}(a^{i+1})}{T(i)})\}\label{eq::samping}
\end{equation}

Notably, we used the empirical definition of $T(i)$, where $T(i) = \frac{T_1}{i + 1}$~\cite{Kirkpatrick1983, Lang2019}.
Experimentally, we set $T_1 = 100$, and the number of iterations $i_{max} = 200$.

\vspace{4px}\noindent{\bf Comparisons of Optimization Performance.}
Fig.~\ref{fig::example_energy_traces} demonstrates an example optimization result while attempting to place a step in front of sink that consist of four anchoring surfaces (Fig.~\ref{fig::importance_map}c).
To better demonstrate the merits, we used greedy approach that tries each possibilities~(Fig.~\ref{fig::example_energy_traces}b) and simulated annealing approach (Fig.~\ref{fig::example_energy_traces}c).
Fig.~\ref{fig::example_energy_traces}a visualizes the $C_{total}$ at each pixel on the anchoring surfaces with greedy approach, where darker area indicating a lower $C_{total}$.
We showed that the greedy approach and simulated annealing approach need to make $2650$ and $49$ attempts respectively before finding the global minimum.

\section{Implementations}\label{sec::implement}
We implemented the authoring pipeline on an iPad ($9$th generation) and the consumption pipeline on the Quest Pro~\cite{QuestPro} due to its colored passthrough, as well as eye~\cite{OculusEyeTracking} and hand tracking~\cite{OculusHandTracking, Chen2022PrecisionDrawing} capabilities.
We also implemented an optimization server using Unity 2021.3.9~\cite{unity20210309} on a separate machine, and a Flask server for managing document (\S\ref{sec::authoring}) and spatial profiles (\S\ref{sec::consuming}).

\section{User Studies}\label{sec::study}

\noindent Two \textit{within-subject} studies were designed to evaluate \sysname.
$12$ participants (PA1 - PA12) were recruited for evaluating the authoring pipeline  and another $12$ participants (PC1 - PC12) were recruited for evaluating consumption pipeline.
During the evaluation, participants either authored or consumed the MR experiences for three cooking tasks (T1 - T3) that could be easily conducted in a typical office kitchen.  
Each study consists of two sessions where participants need to complete the designated tasks that involves with four key objects: {\it microwave}, {\it fridge}, {\it sink}, and {\it countertop}~(Fig.~\ref{fig::study_design}).
We used T1 as the training task, through which participants could get familiar with the designed interfaces.
T2 and T3 were used for formal evaluations, each of which could be completed within $10 \sim 15$~min.
Appendix~\ref{app::tasks} provides the details of instructions.

\subsection{User Study 1: Authoring Pipeline}\label{sec::study::authoring}
\noindent The first study aims to understand how well our authoring pipeline can support authors in easily and rapidly creating an MR instruction experience with manual or ML-assisted mode (\S\ref{sec::authoring::edit}).
We aim to tackle three Research Questions~(RQs): 

\begin{itemize}[leftmargin=*]
    
    \item \textbf{(RQ1)} How the proposed authoring pipeline usable?
    
    \item How the ML could support \textbf{(RQ2)} a {\it faster} and \textbf{(RQ3)} an {\it easier} authoring experience?
    
\end{itemize}

\vspace{4px}
\noindent{\bf Participants and Procedures}.
PA1 - PA12 (age, \mbox{$M = 25.33$}, $SD = 2.81$, \incl~ eight males and four females) were recruited.
Six participants disagreed that they are experts of the designated tasks, with the remaining participants held a neutral opinion.
After training participants to use both of interfaces with T1, participants then completed one task for each interface condition (manual and ML-assisted). 
The tasks and interface conditions were counterbalanced across participants (Fig.~\ref{fig::study_design} in Appendix~\ref{app::study_results}). 
Participants were finally instructed to briefly test the MR experience that was authored, to see if they were satisfied with the authoring outcomes inside MR.
After each session, participants were asked to rate how strongly they agree with three prompts, shown in Fig.~\ref{fig::authoring_eval}e, in a $5$-point Likert scale. 
Participant were then invited to fill out the NASA TLX~\cite{Hart1988NASATLX}, followed by System Usability Scale (SUS)~\cite{Brooke1996sus}, as approximations of perceived workload and level of system usability.
A semi-structured interview was conducted to understand participants' responses.
The study on average took $37.87$~min ($SD = 5.45$~min).

\vspace{4px}
\noindent{\bf Measures.}
We analyzed the differences in the overall SUS, weighted TLX scores, and Task Completion Time~(TCT) while extracting document profiles using both the manual and ML-supported modes.
Shapiro-Wilk test~\cite{Shapiro1965} was used to check the normality of data in each catalogue.
Repeated Measure Analysis of variance (RM-ANOVA)~\cite{Girden1992} with Tukey's HSD~\cite{Abdi2010} ($\alpha = 0.05$) were used for analyzing statistical significance and \posthoc comparisons.
Upon a failure of normality check, the Aligned Rank Transform~(ART)~\cite{Wobbrock2011}, followed by ART contrast test~(ART-C)~\cite{Elkin2021} were used as the nonparametric approach for statistical significance analysis and \posthoc test\footnote{Notations for indicating the \posthoc test results shown in Fig.~\ref{fig::authoring_eval}, \ref{fig::consumption::evaluation_measure}, and \ref{fig::consumption::overall}: $*$ ($p < .05$), $**$ ($p < .01$), $***$ ($p<.001$).}.
The partial eta square ($\eta_p^2$) was used to evaluate the effect size for ART and RM-ANOVA, with $.01$, $.06$ and $.14$ indicating the thresholds for small, medium and large effect size~\cite{Cohen1998}.
We used thematic analysis~\cite{Braun2012}, and deductive and inductive coding~\cite{Elo2008} to analyze qualitative data, to understand participants' experience.

\vspace{4px}\noindent
{\bf Results and Discussions}

\noindent
Overall, most participants found the authoring pipeline easy to use (RQ1), and agreed that the ML-assisted mode could help the authoring experience faster (RQ2) and easier (RQ3).

\vspace{4px}
\noindent{\bf Is the Proposed Authoring Pipeline Usable (RQ1)?}
Both manual ($M = 70.61$, $SD = 17.23$) and ML-assisted ($M = 86.88$, $SD = 8.33$) modes demonstrate good system usability, where an overall SUS score $\geq$ $68$ is interpreted as ``good''~\cite{Sauro2011, Sauro2016}.
All participants agreed that both pipelines are easy to use (Fig.~\ref{fig::authoring_eval}a). 
For example:
{\it ``they are easy to use! [... and] integrated very good!''}~(PA1) and {\it ``it has a really good workflow''}~(PA9).
While $11$ participant believed that they could author such MR experience with both interfaces rapidly (Fig.~\ref{fig::authoring_eval}e), PA10 held a neutral opinion, as she expected a fully automated system to extract the key objects associated with each instruction step.
Overall, $10$ participants were satisfied with the MR experience they authored (Fig.~\ref{fig::authoring_eval}e).
Examples include:
{\it ``a cool way to transfer knowledge''}~(PA2) and {\it ``useful to see [the instruction] while cooking''}~(PA12).
Few participants suggested the reason(s) for being not fully satisfied with the authored MR experience.
For example,
{\it``I think I made some mistake when I author it. So it guided me to the wrong spot''}~(PA5)
and {\it ``I have to do [the tasks with authored experience] by myself, to see what it is like for me to experience that first before having a novice do it so''}~(PA12).
These implied the lack of ways to enable the authors to revise the authored experience inside MR iteratively.

\begin{figure}[t]
    \centering
    \includegraphics[width=0.48\textwidth]{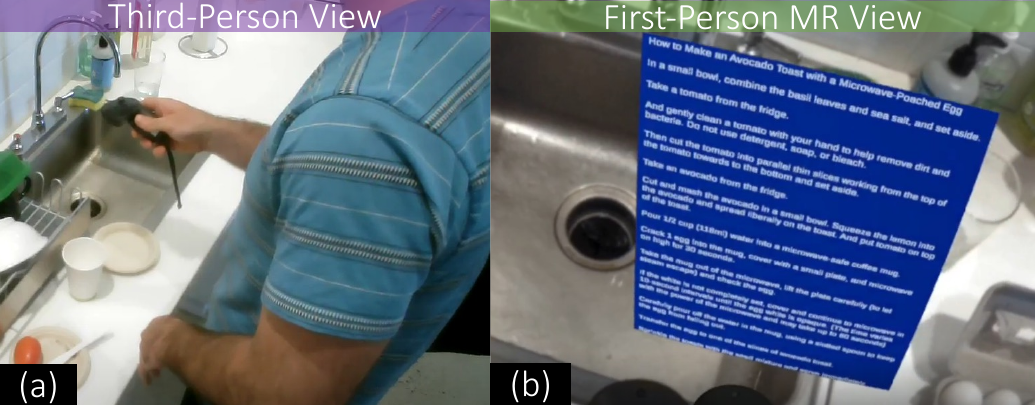}
    \vspace{-0.3in}
    \caption{Baseline scene. Third-person (a) and first-person view through MR (b).}
    \Description{Figure 12.a shows the third-person view of an example baseline scene, in which a participant was hold a touch controller using right hand; Figure 12.b describes the first-person view of Figure 12.a, in which a virtual monolithic document is shown up through MR, and attached to the right touch controller.}
    \vspace{-0.1in}
    \label{fig::consumption::baseline}
\end{figure}

\vspace{4px}\noindent{\bf Can ML Support a Faster Authoring Experience~(RQ2)?}
The RM-ANOVA showed a reduced overall TCT ($F_{1, 22} = 16.66$, $p < .001$, $\eta_p^2 = .43$, Fig.~\ref{fig::authoring_eval}c) and average TCT (ART: $F_{1, 22} = 35.98$, $p < .001$, $\eta_p^2 = .77$, Fig.~\ref{fig::authoring_eval}d) for authoring each step while using ML-assisted mode, versus manual approach.
Most participants echo such observations.
For example:
{\it ``it could help me saving time, because [the key objects have] already [been] filled out [...] it speeds up by maybe a half a second [for each instruction]''} (PA3)
and {\it ``it helps me to save a lot of time! And that makes it a lot more convenient!''}~(PA4)

\vspace{4px}\noindent{\bf Can ML Support an Easier Authoring Experience (RQ3)?}
The RM-ANOVA demonstrated a higher SUS ($F_{1, 22} = 8.65$, $p = .008$, $\eta_p^2 = .28$, Fig.~\ref{fig::authoring_eval}a) and a lower TLX score (ART: $F_{1, 22} = 17.24$, $p = .002$, $\eta_p^2 = .61$, Fig.~\ref{fig::authoring_eval}b) of the ML-assisted mode, compared to the manual counterpart.
Most participants appreciated the convenience and helpfulness brought by the predicted key objects.
First, nearly all participants believed that the ML helped on reducing effort for tagging key objects.
For example: 
{\it ``ML brought less effort, I just need to check if the predicted key object is correct or not [...] Even if I still need to check it, I don't have to pay $100$\% of attention. I don't have to do all the thing. I just have to do part of the thing''}~(PA3), 
Particularly, the features of real-time predicting the new key object while modifying a specific step were favored by some participants.
For example:
{\it ``when I saw one of the step to be very long and [are associated with two key objects] [...] capable of predicting key object after being modified is obviously helpful! And also the opposite feature where you could just like combine two tasks, followed by generating predicted key object! [...] it gives more flexibility while segmenting the instruction step''}~(PA10).

Second, some participants highlighted the helpfulness for the mental thought process for ML-assisted mode. 
{\it ``I was able to create a mental map of how I will spatially move across at different instances [by looking at the predicted key objects]''}~(PA4)
and {\it ``it could help me make a decision''}~(PA3).
Particularly, PA5 appraised the feature of predicting key object in real-time while revise the instruction: {\it ``[while adding or editing the steps based on existing paper instruction], the real-time predicted location could help create a more clear instruction step. For example, if I type `heat the water', and the predicted location is oven for somehow, then I might just type `heat the water in the microwave' to make the instruction more clear''}
However, PA10 held an opposite opinion: {\it ``I just read the instruction step and then check if this assigned [key object] was countertop or not, and changed it to countertop rather than going and checking the other options. [...] but I did not read [the predicted key object] first.'' }

Finally, few participants suggested the merits of using color scale to visualize the confidence of predicted key object.
For example: 
{\it ``color could be helpful for conveying uncertainty''}~(PA9)
and {\it``color confidence is important to me. If it's red, I would be more aware of checking whether this is actually correct or not''}~(PA12).

\subsection{User Study 2: Consumption Pipeline}\label{sec::study::consuming}
The second study aims to evaluate the consumption pipeline and attempts to address:
{\it ``how the \sysname~could help the consumers to complete the designated activities {\it faster} and {\it easier}?''}

\begin{figure}[t]
    \centering
    \includegraphics[width=0.48\textwidth]{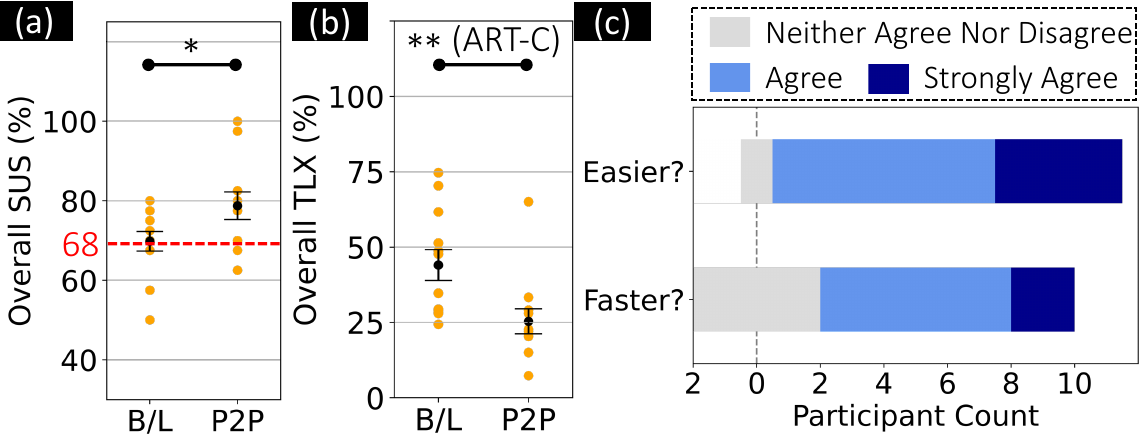}
    \vspace{-0.3in}
    \caption{Consumption pipeline evaluation results. (a - b) Overall SUS and TLX score; (c) Survey results of how participants considered the overall consumption experience of \sysname~faster and easier, versus baseline. ``B/L'' and ``P2P'' indicate baseline and \sysname~condition.}
    \Description{Figure 13.a shows the overall SUS score of baseline and PaperToPlace. ``*'' was annotated to indicate the statistical significance results; Figure 13.b describes the overall TLX score of baseline and PaperToPlace. ``** (ART-C)'' was annotated to indicate the statistical significance results; Figure 13.c shows the survey results of consumption pipeline evaluations. One, seven, and four participants held a ``neither agree nor disagree'', ``agree'', and ``strongly agree'' opinions toward ``easier?'' prompt. Four, six, and two participants held a ``neither agree nor disagree'', ``agree'', and ``strongly agree'' opinions toward ``faster?'' prompt.}
    \vspace{-0.10in}
    \label{fig::consumption::overall}
\end{figure}

\begin{figure*}[t]
    \centering
    \includegraphics[width=\textwidth]{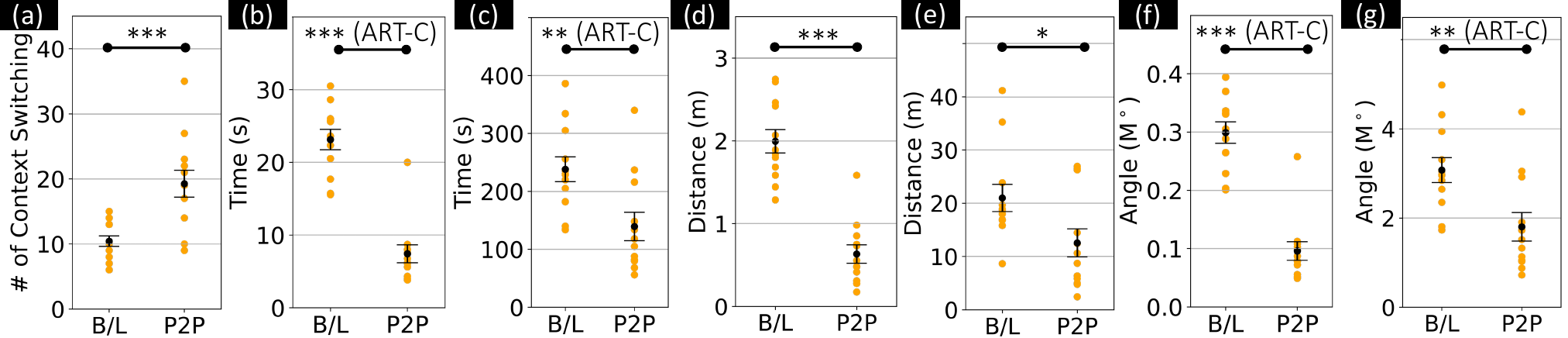}
    \vspace{-0.3in}
    \caption{Results of context switching evaluations. (a) The total number of context switching while using the monolithic document and \sysname. The average time (b), $d_{head}$ (d), and $\theta_{head}$ (f) during each episode; The total time (c), $l_{head}$ (e), and $\theta_{head}$ (g) during all episode while completing the task. ``B/L'' and ``P2P'' indicate the baseline and \sysname~conditions.}
    \Description{Figure 14.a shows the number of context switching for baseline and PaperToPlace system. ``***'' was annotated to indicate the statistical significance results; Figure 14.b describes the average time for each episode for baseline and PaperToPlace system. ``*** (ART-C)'' was annotated to indicate the statistical significance results; Figure 14.c demonstrates the total time for all episodes for baseline and PaperToPlace system. ``** (ART-C)'' was annotated to indicate the statistical significance results; Figure 14.d describes the average total path of the head movement of each episode. ``***'' was annotated to indicate the statistical significance results; Figure 14.e shows the total path of the head movement of each episode. ``*'' was annotated to indicate the statistical significance results; Figure 14.f shows the average angular change of head rotation of each episode. ``*** (ART-C)'' was annotated to indicate the statistical significance; Figure 14.g demonstrates the total angular change of head rotation of all episodes. ``** (ART-C)'' was annotated to indicate the statistical significance.}
    \vspace{-0.15in}
    \label{fig::consumption::evaluation_measure}
\end{figure*}

\vspace{4px}\noindent{\bf Participants and Procedures}.
PC1 - PC12 (age, $M = 27.83$, $SD = 6.55$, \incl~ six males and six females) were recruited.
All participants were {\it not} familiar with the designated tasks.
We also built a baseline experience where the consumers could read existing monolithic instruction document inside MR (Fig.~\ref{fig::consumption::baseline}).
Instead of asking participants to read a paper document, rendering a virtual monolithic document attached to the touch controller could minimize the impacts of confounding factors caused by uncomfort of the headset. 
We designed two sessions (Fig.~\ref{fig::study_design}) with counterbalancing being considered to minimize the impacts of prior learning experience and task familiarity.
Before each session, T1 was used to help participants learn and familiarize with the system.
During the session, participants were instructed to complete \mbox{T2/T3} using the baseline and \sysname.
Participants were invited to fill out NASA TLX~\cite{Hart1988NASATLX} and SUS~\cite{Brooke1996sus} at the end of each session, followed by a semi-structured interview.

\vspace{4px}\noindent{\bf Measures. }
To understand the performance of context switching, we defined each {\bf episode} as {\it the interval between the time when participants stopped a task to seek instructions and when they returned to the task}.
During each episode, we analyzed the \textbf{(i)} time; \textbf{(ii)} the distance of path of head movement ($d_{head}$); \textbf{(iii)} the angular changes of the forward direction of the head ($\theta_{head}$).
The data lies outside of such intervals are out of our scope, as the performance of real-world activities could be affected by participants' prior cooking experience.
Same approaches in \S~\ref{sec::study::authoring} were used to analyze the questionnaire responses and participant's qualitative feedback.
The study on average took $59.40$~min~($SD = 5.80$~min).

\vspace{4px}\noindent{\bf Results and Discussions. }
Overall, most participants believed that the \sysname~could help the consumers to complete the designated tasks faster and easier (Fig.~\ref{fig::consumption::overall}c). 
Quantitatively, we demonstrated a higher overall SUS score ($F_{1, 22} = 4.44$, $p = .046$, $\eta_p^2 = .17$, Fig.~\ref{fig::consumption::overall}a) and a lower perceived workloads (ART: $F_{1, 22} = 18.52$, $p = 0.001$, $\eta_p^2 = .63$, Fig.~\ref{fig::consumption::overall}b).
Based on participants' feedback, we now discuss how \sysname~could help participants complete the designated tasks faster and easier.

\vspace{4px}\noindent{\bf Context Awareness Reduces the Effort of Context-Switching.}
\sysname~reduces the average time (ART: $F_{1, 22} = 49.18$, $p < .001$ $\eta_p^2 = .82$, Fig.~\ref{fig::consumption::evaluation_measure}b), $d_{head}$ ($F_{1, 22} = 57.29$, $p < .001$ $\eta_p^2 = .72$, Fig.~\ref{fig::consumption::evaluation_measure}d) and $\theta_{head}$ (ART: $F_{1, 22} = 51.19$, $p < .001$ $\eta_p^2 = .82$, Fig.~\ref{fig::consumption::evaluation_measure}f) on each episode.
While \sysname~on average leads to frequent document readings ($F_{1, 22} = 15.77$, $p < .001$, $\eta_p^2 = .42$, Fig.~\ref{fig::consumption::evaluation_measure}a), the accumulated time (ART: $F_{1, 22} = 9.20$, $p = .020$ $\eta_p^2 = .42$, Fig.~\ref{fig::consumption::evaluation_measure}c), $d_{head}$ ($F_{1, 22} = 5.30$, $p = .030$ $\eta_p^2 = .19$, Fig.~\ref{fig::consumption::evaluation_measure}e) and $\theta_{head}$ (ART: $F_{1, 22} = 7.48$, $p = .019$ $\eta_p^2 = .40$, Fig.~\ref{fig::consumption::evaluation_measure}g) of all episodes are reduced.

First, participants suggested the convenience for referring back to the instructions repeatedly with \sysname.
For example: {\it ``[with baseline], I need to check it back and forth every time while trying to grab food from the fridge [...] [\sysname] gives me feeling like [the instruction] is just on my side. It's like always on my side, like right beside my head''}~(PC9) and 
{\it ``because [the step] would always be right there with just a little bit of information I need, I think it'd be very useful''}~(PC11).
Most participants explicitly highlighted the benefits of finding optimal placements, without causing occlusions to the key interaction areas.
For example: {\it ``it is useful to bring the instruction step to me by just a pinch.''}~(PC10), {\it ``I like the position of instruction step!''}~(PC4), and {\it ``I think it is useful! And especially the function of where you pinch again, it will move to another location, so it can ensure [the step] will never block your sight''}~(PC8).
More example of instruction step placements could be referred to \mbox{Fig.~\ref{fig::consumption::examples}e - h}.
However, PC6, who unveiled her ADHD~\cite{adhd}, suggested: 
{\it ``I was distracted with [the \sysname], because [when the step occasionally was not anchored on the optimal position] the information was here and I was there. It reminded me of the moments that I forgot what I was supposed to do, or what I have to do.''}~

Second, most participants acknowledged the helpfulness by establishing connections between instructions and key objects.
For example: {\it ``I like how it took me to the sink because this activity has to be near the sink. That's a very helpful on spatial understanding!''}~(PC4), {\it ``Although we know where the fridge is, having that is really convenient to just not give anything a thought and do things as per the instructions''}~(PC10).

Finally, five participants also mentioned the merits of hands-free of \sysname, compared to the baseline where the consumers need to hold the virtual document.
For example {\it ``[baseline] is more cumbersome, because I need to free one hand and make the hand very clean to make sure that the hand is clean to touch the controller''}~(PC5).

\vspace{4px}\noindent{\bf Segmented Instruction Helps Findings the Relevant Information Easier.}
All participants suggested that the segmented instructions is helpful, \eg~{\it ``[instructions] need to be as concise and as short as possible to be read at the same time. [\sysname] did its job!''} (PC5).
Most participants suggested the merits of reducing stress while translating instruction into real world activities.
For example: {\it ``I have more calmness [with \sysname], because [with baseline] I was seeing everything all at once, and that was giving me the feeling I'm in a hurry.''} (PC7),
and {\it ``looking at the entire document at once was so hard that I forgot where I have to keep following from start to finish to find where I was. But [\sysname] gave me one by one instructions, which is super easy!''}~(PC3)

\section{Limitations and Future Works}\label{sec::future}

\noindent We identify our limitations from four perspectives.

\vspace{4px}
\noindent{\bf (1) Enabling an Iterative Authoring Process.}
We observed that during authoring, participants intended to segment the instruction steps by {\it only} considering whether only one key object is associated with the specific step, without synthesizing other factors (\eg~the density of the information contained by single step while viewed inside MR).
However, this cannot ensure a satisfied MR experience from the perspective of the consumers.
Future work might investigate potential {\it iterative} authoring workflow that allows the authors to refine their authored document profile inside MR while piloting the created instructional MR experience.

\vspace{4px}\noindent{\bf (2) Transforming Richer Metadata into MR Experiences.}
We consider that the metadata of each instruction step {\it only} contains the text of the step and the key objects that the virtual step should be anchored on (\S\ref{sec::system_overview::design_insights}).
This might not be realistic for real-world instruction documents with heterogeneous kinds of metadata, such as the duration information, the caveats that usually requires the consumers' attention, and the notifications from the environmental sensors.
For example: {\it ``I would like to have warning text, like `do not use detergent', maybe show up in a different color or something''}~(PC4).
Future research might investigate the richer metadata that need to be augmented inside the MR experience, and the methods to use existing language models to extract such metadata as well as transform them into spatialized and context-aware MR experience.

\vspace{4px}
\noindent{\bf (3) Automatically Switching between Instruction Steps and Triggering the Position Update of the Instruction Label.}
We currently required consumers to explicitly click the virtual button to switch to the next instruction step, and to pinch to update the current position of the instruction step on demand~(\S\ref{sec::consuming::interactions}).
While participants (\eg~PC10) with some prior MR experience felt it is {\it ``easy and useful''} to use the virtual hand menu and pinch gesture, others (\eg~PC1) suggested the frustrations of occasional failures of pinch gesture detections and the virtual button clicking.
Future work might consider designing a state machine, which could specify how to switch to the subsequent step {\it automatically} based on user's activities that might be inferred from face (\eg~\cite{Chen2021EXGSense}), body (\eg~\cite{OculusBodyTracking}) and environmental (\eg~\cite{Boovaraghavan2023, Agarwal2019virtual, Chen2020Captag}) sensor data. 

\vspace{4px}\noindent{\bf (4) Supporting a Broader Range of Applications. }
While many participants believed {\it ``cooking demonstrates [\sysname] very well''}~(PC4), we only evaluated on cooking instructions, due to the poor quality of passthrough capabilities of Quest Pro~\cite{QuestPro}; availability of dataset to fine tune language model for alternative instruction activities; and limited study resources.
Future work might explore other activities with more powerful language model such as GPT and prompt engineering techniques being used for creating document profiles. 
Participants also emphasized the values of adaptive placements (\eg~{\it ``the adaptive placement of instruction is definitely useful for paper cutting! I don't want to cut my hand. And I wanted the instruction to be always besides my hand''}~(PC9)) and reduced context switching (\eg~{\it ``in the gym, where I need some instructions to teach me how to use the equipment, such as how you hold the gears with good postures. [...] with [baseline], it is less efficient and [I have] to stop in between and read the instructions''}~(PC2)) that might be transferred to other activities.
Another direction is to investigate the support for finer grained tasks that might be involved with moving objects, leading to a dynamically changed spatial profile.
For example, {\it``if you are doing PCB soldering, it might be hard to track that tiny component and to pinpoint the exact location on the PCB board. But if [\sysname] can do that, it will be super helpful!''}~(PC8).
This requires high quality passthrough and capabilities to track real-time location of the electronic components which are considered as {\it non}-static key objects.
Instead of using Quest Pro, future researchers might consider a more recent higher-end headset, \eg~Vision Pro~\cite{VisionPro}.
\section{Conclusion}\label{sec::conclusions}
We present and evaluate \sysname, comprising {\it an authoring pipeline}, which allows authors to rapidly transform existing paper instructions into a MR experience, and {\it a consumption pipeline}, which enables consumers to view spatialized instructions using a context-aware approach.
Two within-subject studies with two different cohorts of $12$~participants demonstrate the usability and effectiveness of the proposed authoring and consumption \mbox{workflows}.

\begin{acks}
    We thank the insightful feedback from the anonymous reviewers. 
    We appreciate the discussions with fellow researchers from Adobe Research, including Mira Dontcheva, Alexa Sui, Stephen DiVerdi, Ryan A. Rossi, Chang Xiao, Geonsun Lee, Mustafa Do\u{g}a Do\u{g}an, and Shakiba Davari.
    We thank Matin Yarmand for the help on the narration of the accompanion video.
\end{acks}

\bibliographystyle{ACM-Reference-Format}
\bibliography{reference}

\newpage
\appendix
\section{Experimental Tasks}\label{app::tasks}
The selected tasks for final user study (\S\ref{sec::study}) include:

\begin{itemize}[leftmargin=*]

\item \textbf{(T1)} Microwave Scrambled Eggs~\cite{ScrambleEggs};

\item \textbf{(T2)} Quick Microwave-Poached Eggs on Avocado Toast~\cite{AvocadoToast};

\item \textbf{(T3)} Instant Mac `n' Cheese~\cite{MacNCheese}; 

\end{itemize}

Table~\ref{table::task1}, Table~\ref{table::task2} and Table~\ref{table::task3} offers the supplementary material regarding the specific instruction steps of the experimental tasks T1, T2, and T3 respectively.
Notably, T1 was used as the training tasks for participants to learn and get familiar with the interfaces (Table~\ref{table::task1}).
Assuming the results from OCR is fully correct (\ie~all texts of all instruction steps could be successfully extracted), the overall accuracies of the associated key objects predicted by our pre-trained language model for each experimental tasks are $50\%$~(T1), $84.62\%$~(T2), and $80.00\%$~(T3).
Notably, the overall accuracies for T2 and T3 are closed to our benchmark results while fine-tuning the BERT model in \S\ref{sec::authoring::edit}, which is $82.13\%$.
This ensures the results yielded by the evaluation of authoring pipeline (\S\ref{sec::study::authoring}) is generalizable to some extend.
Although the overall accuracy of T1 is far lower than our benchmark results due to the relative short of instruction document, the instruction of T1 was only used for participants to familiarize themselves with the given interfaces (either on iPad or inside MR), and the data yielded by T1 was excluded from our evaluation results in \S\ref{sec::study}.

\begin{table}
    \small
    \centering
    \begin{tabular}{l} 
     \hline
     \textbf{Step Descriptions} \\
     \hline\hline 

     \makecell[l]{$\bullet$~Spray microwave-safe container (\eg~mug, ramekin, or egg cooker) \\ with cooking spray or wipe lightly with vegetable oil.} \\ 
    
     \makecell[l]{$\bullet$~Whisk eggs, milk, salt and pepper in container (or whisk ingredients \\ in another bowl and pour into microwave container). If using a mug \\ or ramekin, cover with plastic wrap, pulling back small area for \\ venting. If using an egg cooker, place lid on cooker base, lining up \\ notches. Twist to secure.}  \\
     
     \makecell[l]{$\bullet$~Microwave on Medium-High ($70$\% power) for 90 seconds, stirring \\ several times during cooking.} \\
     
     \makecell[l]{$\bullet$~Cover and let stand for 30 seconds to 1 minute before serving. Eggs \\ will look slightly moist, but will finish cooking upon standing.} \\
     
     \hline
    \end{tabular}
    \caption{Experimental cooking recipe for making basic microwave scrambled eggs (T1).}
    \label{table::task1}
\end{table}

\begin{table}
    \small
    \centering
    \begin{tabular}{l} 
     \hline
     \textbf{Step Descriptions} \\
     \hline\hline 

     \makecell[l]{$\bullet$~In a small bowl, combine the basil leaves and sea salt, and set aside.} \\ 
    
     \makecell[l]{$\bullet$~Take a tomato from the fridge.}  \\
     
     \makecell[l]{$\bullet$~And gently clean a tomato with your hand to help remove dirt and \\ bacteria. Do not use detergent, soap, or bleach.} \\
     
     \makecell[l]{$\bullet$~Then cut the tomato into parallel thin slices working from the top \\ of the tomato towards to the bottom and set aside.} \\
     
     \makecell[l]{$\bullet$~Take an avocado from the fridge.} \\
     
     \makecell[l]{$\bullet$~Cut and mash the avocado in a small bowl. Squeeze the lemon into \\ the avocado and spread liberally on the toast. And put tomato on top \\ of the toast.} \\
     
     \makecell[l]{$\bullet$~Pour $1/2$ cup ($118$~ml) water into a microwave-safe coffee mug. } \\
     
     \makecell[l]{$\bullet$~Crack 1 egg into the mug, cover with a small plate, and microwave on \\ high for 30 seconds. } \\
     
     \makecell[l]{$\bullet$~Take the mug out of the microwave, lift the plate carefully (to let \\ steam escape) and check the egg. } \\
     
     \makecell[l]{$\bullet$~If the white is not completely set, cover and continue to microwave in \\ $10$-second intervals until the egg white is opaque. (The time varies with \\ the power of the microwave and may take up to 60 seconds).} \\
     
     \makecell[l]{$\bullet$~Carefully pour off the water in the mug, using a slotted spoon to keep \\ the egg from falling out. } \\
     
     \makecell[l]{$\bullet$~Transfer the egg to one of the slices of avocado toast. } \\
     
     \makecell[l]{$\bullet$~Sprinkle the toasts with the seed mixture and serve immediately.} \\
     
     \hline
    \end{tabular}
    \caption{Experimental cooking recipe for making an avocado toast with a microwave-poached egg (T2).}
    \label{table::task2}
\end{table}

\begin{table}
    \small
    \centering
    \begin{tabular}{l} 
     \hline
     \textbf{Step Descriptions} \\
     \hline\hline 

     \makecell[l]{$\bullet$~Find a mug that holds twice the volume of your dry pasta – the \\ bigger, the better.} \\ 
    
     \makecell[l]{$\bullet$~Add the macaroni.}  \\
     
     \makecell[l]{$\bullet$~Add some water.} \\
     
     \makecell[l]{$\bullet$~Cover with cling film and pierce $3$ times.} \\
     
     \makecell[l]{$\bullet$~Stand the mug in a microwave-proof bowl to catch any spillages, and \\ cook in the microwave on high for $2$ minutes. The liquid will bubble \\ up and over the sides, so tip any liquid from the bowl back into \\ the mug (be careful as it will be very hot) and give it a good stir.  } \\
     
     \makecell[l]{$\bullet$~Leave to stand for $1$ minute.} \\
     
     \makecell[l]{$\bullet$~Repeat twice more or until the pasta is cooked (it may take longer \\ depending on the pasta).} \\
     
     \makecell[l]{$\bullet$~Then remove from the microwave. } \\
     
     \makecell[l]{$\bullet$~Repeat twice more or until the pasta is cooked (it may take longer \\ depending on the pasta).} \\
     
     \makecell[l]{$\bullet$~Stir through the butter, cheese and spinach or Marmite, if using. } \\
     
     \makecell[l]{$\bullet$~The heat from the pasta should melt the cheese and wilt the spinach, \\ but if not, pop back in the microwave for $30$ seconds.} \\
     
     \hline
    \end{tabular}
    \caption{Experimental cooking recipe for making microwaved mac `n' cheese (T3).}
    \label{table::task3}
\end{table}

\section{User Study Results}\label{app::study_results}
\begin{figure}[h!]
    \centering
    \includegraphics[width=0.48\textwidth]{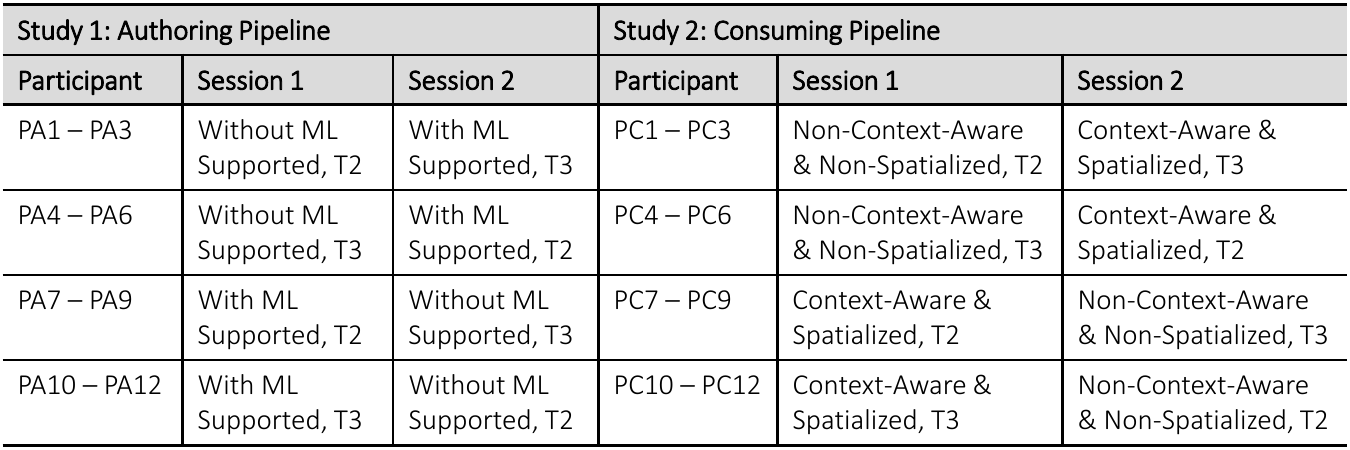}
    \vspace{-0.25in}
    \caption{Study design for evaluating \sysname. Each participant needs to conduct session 1 and session 2 in order. T2 and T3 were used for formal evaluation while T1 was used for training purposes.}
    \Description{Figure 15 shows the study design for evaluating PaperToPlace. Each participant needs to conduct session 1 and session 2 in order. T2 and T3 were used for formal evaluation while T1 was used for training purposes.}
    \label{fig::study_design}
\end{figure}

This section presents supplementary material for \S\ref{sec::study}.
Fig.~\ref{fig::study_design} shows the specific tasks and interface conditions that were assigned while evaluating authoring and consumption pipeline.
Notably, T1 was used for training purposes.

We also provide visualizations of survey responses for authoring pipeline evaluations (\S\ref{sec::app::study_results::authoring}) as well as consumption evaluations (\S\ref{sec::app::study_results::consuming}).

\begin{figure}[t]
    \centering
    \includegraphics[width=0.46\textwidth]{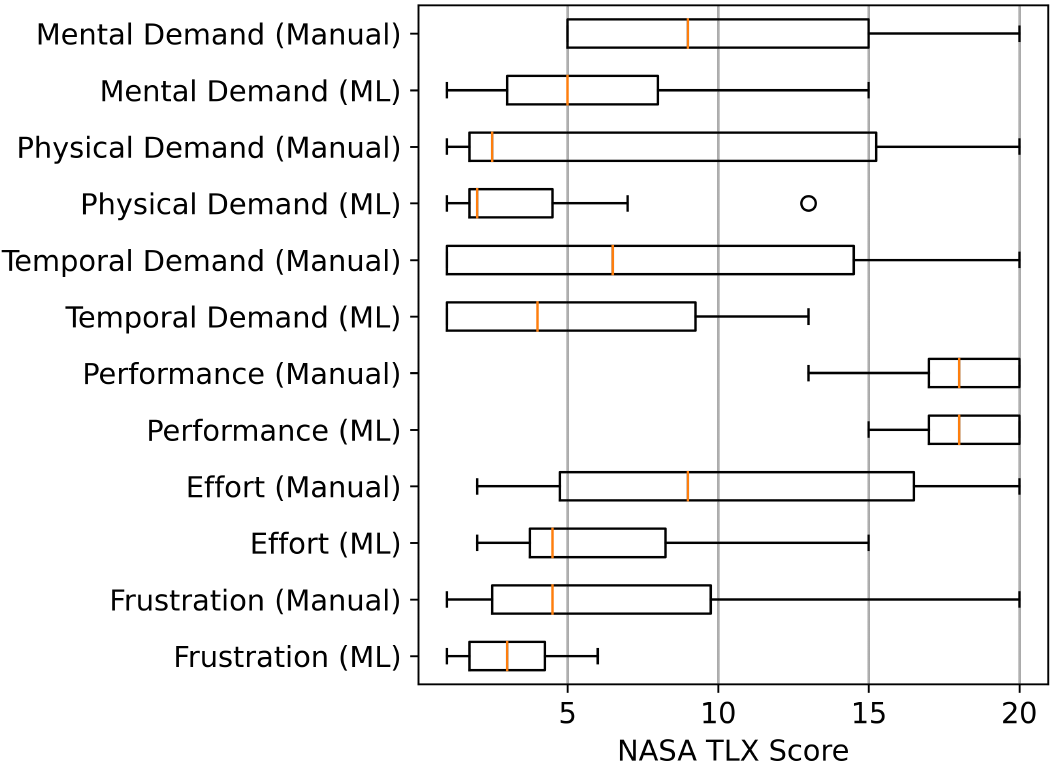}
    \vspace{-0.15in}
    \caption{Survey results of NASA TLX questionnaires. We use ``Manual'' and "ML" to indicate the interface condition while manual and ML supported approaches are used for extracting associated key objects from each instruction steps, respectively}
    \Description{Figure 16 shows the survey results of NASA TLX questionnaires.We use ``Manual'' and ``ML'' to indicate the interface condition while manual and ML supported approaches are used for extracting associated key objects from each instruction steps, respectively.}
    \label{fig::appenx::authoring_results::tlx}
\end{figure}

\begin{figure}[t]
    \centering
    \includegraphics[width=0.46\textwidth]{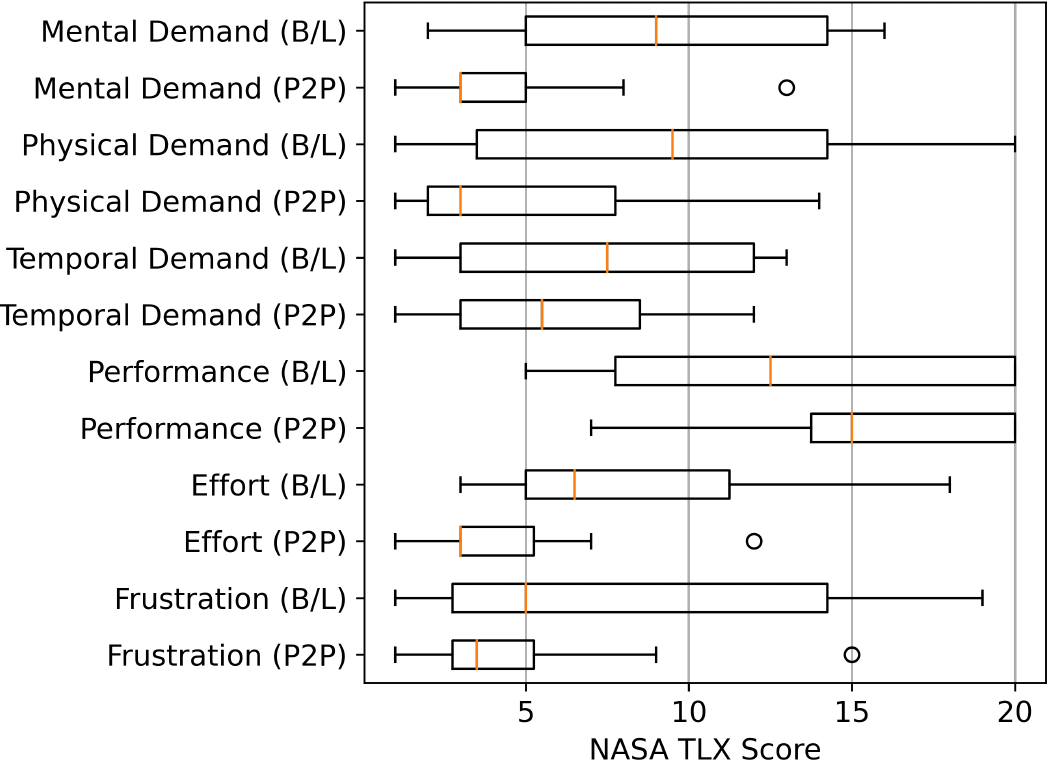}
    \vspace{-0.15in}
    \caption{Survey results of NASA TLX questionnaires. We use ``B/L'' to refer to the baseline interface, and ``P2P'' to indicate \sysname, which delivers spatialized and context-aware instruction step.}
    \Description{Figure 17 shows the survey results of NASA TLX questionnaires.We use ``B/L'' to refer to the baseline interface, and ``P2P'' to indicate PaperToPlace, which delivers spatialized and context aware instruction steps;}
    \label{fig::appenx::consumption_results::tlx}
\end{figure}

\begin{figure*}
    \centering
    \includegraphics[width=0.85\textwidth]{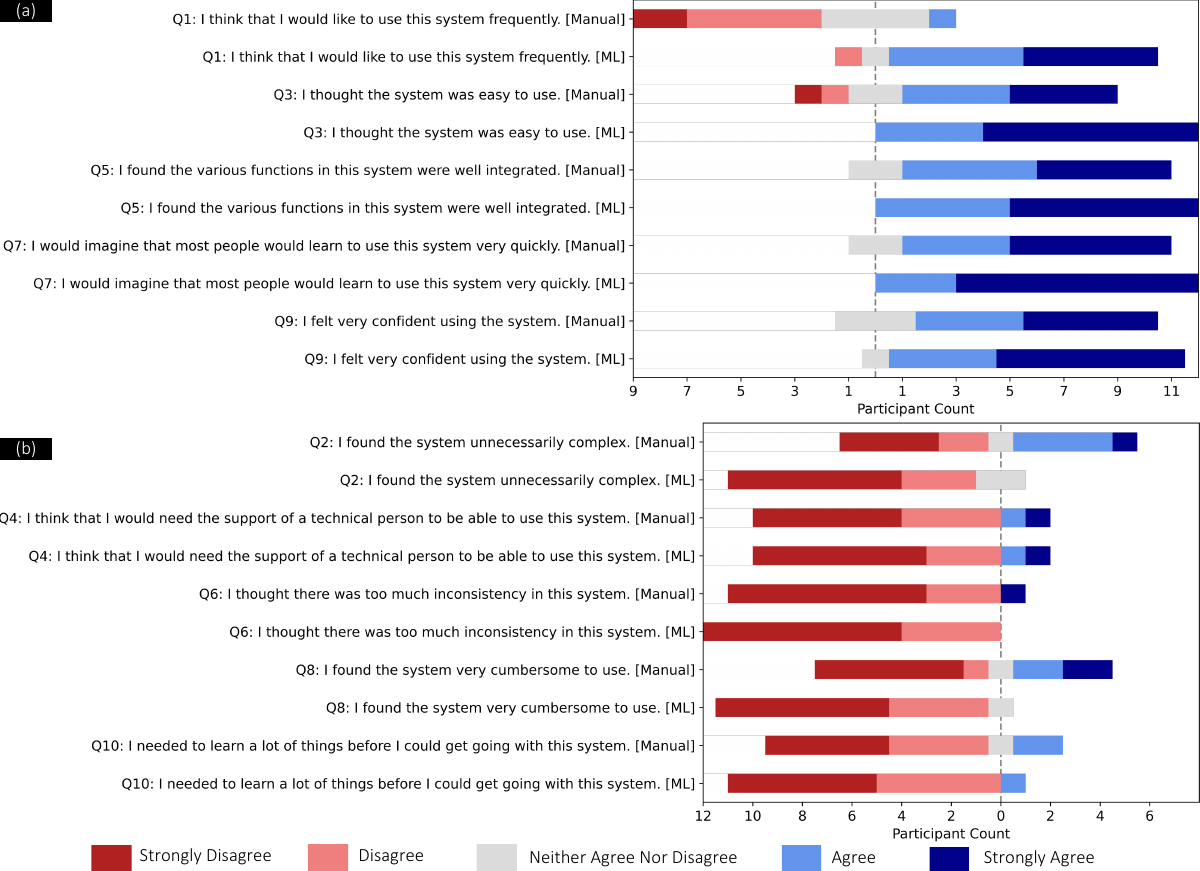}
    \vspace{-0.10in}
    \caption{Survey results of SUS questionnaires of authoring pipeline evaluations. We use ``Manual'' and ``ML'' to indicate the interface condition while manual and ML support approaches are used for extracting associated key objects from each instruction steps, respectively. To increase readability, we cluster the survey results of positive statements~(Q1, Q3, Q5, Q7, Q9) into subplot~(a), where a higher level of agreement indicates a better user experience. The survey results of negative statements~(Q2, Q4, Q6, Q8, Q10) are clustered into subplot~(b), where a lower level of agreement indicates a better user experience.}
    \Description{Figure 18 shows the survey results of SUS questionnaires of authoring pipeline evaluations. We use ``Manual'' and ``ML'' to indicate the interface condition while manual and ML support approaches are used for extracting associated key objects from each instruction steps, respectively.}    
    \label{fig::appenx::authoring_results::sus}
\end{figure*}

\begin{figure*}
    \centering
    \includegraphics[width=0.85\textwidth]{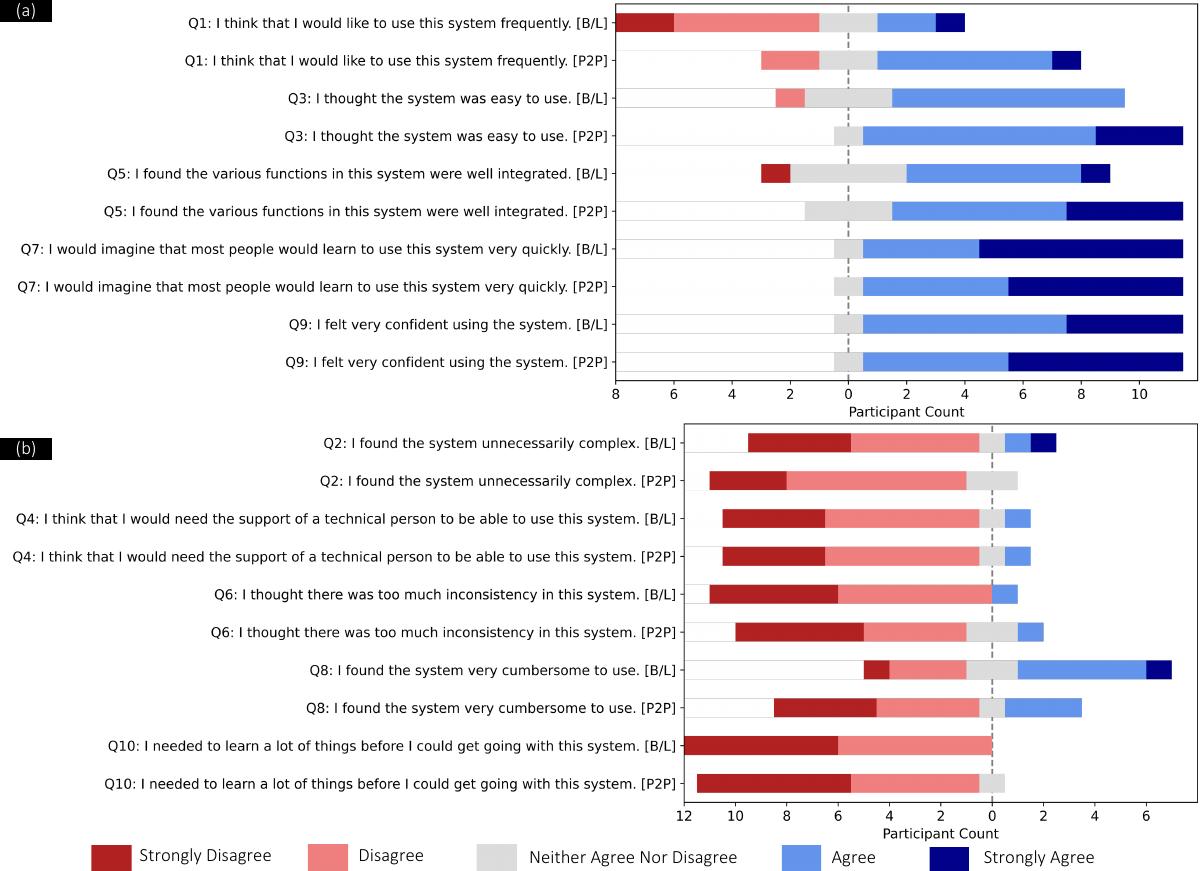}
    \vspace{-0.10in}
    \caption{Survey results of SUS questionnaires of consumption pipeline evaluations. We use ``B/L'' to refer to the baseline interface, and ``P2P'' to indicate \sysname, which delivers spatialized and context-aware instruction step. To increase readability, we cluster the survey results of positive statements~(Q1, Q3, Q5, Q7, Q9) into subplot~(a), where a higher level of agreement indicates a better user experience. The survey results of negative statements~(Q2, Q4, Q6, Q8, Q10) are clustered into subplot~(b), where a lower level of agreement indicates a better user experience.}
    \Description{Figure 19 describes the survey results of SUS questionnaires of consumption pipeline evaluations. We use ``B/L'' to refer to the baseline interface, and ``P2P'' to indicate PaperToPlace, which delivers spatialized and context-aware instruction steps.}
    \label{fig::appenx::consumption_results::sus}
\end{figure*}

\begin{figure*}
    \centering
    \includegraphics[width=\textwidth]{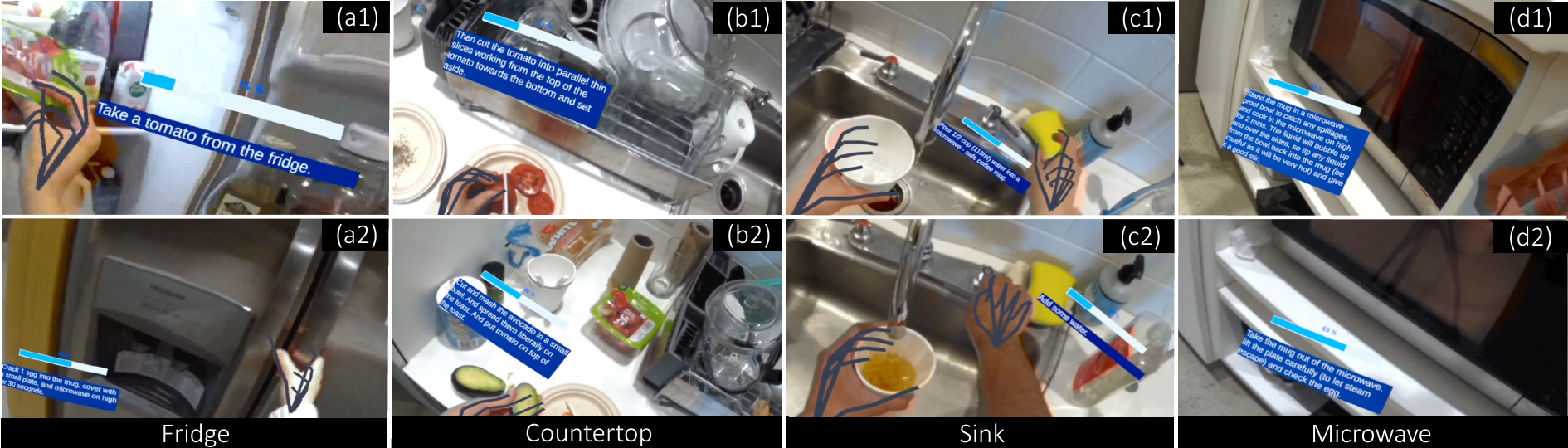}
    \vspace{-0.20in}
    \caption{First-person view through MR of the examples of placing instruction step next to the key objects with our consumption pipeline. Example key objects include fridge (a), countertop (b), sink (c), and microwave (d).}
    \Description{Figure 20.a shows the first-person view through MR of the examples of placing instruction steps next to the fridge with our consumption pipeline; Figure 20.b describes the first-person view through MR of the examples of placing instruction steps next to the countertop with our consumption pipeline; Figure 20.c shows the first-person view through MR of the examples of placing instruction steps next to the sink with our consumption pipeline; Figure 20.d describes the first-person view through MR of the examples of placing instruction steps next to the microwave with our consumption pipeline.}
    \label{fig::consumption::examples}
\end{figure*}

\subsection{Evaluations of Authoring Pipeline}\label{sec::app::study_results::authoring}
Fig.~\ref{fig::appenx::authoring_results::tlx} demonstrates the NASA TLX responses of each perceived workloads from participants PA1 - PA12, while using manual and ML-assisted interfaces to extract document profile from designated paper instruction.
Fig.~\ref{fig::appenx::authoring_results::sus} provides supplementary material of survey results of SUS questionnaires.
For Fig.~\ref{fig::appenx::authoring_results::tlx} and Fig.~\ref{fig::appenx::authoring_results::sus}, we use ``Manual'' and ``ML'' to indicate the interface condition while manual and ML supported approaches are used while extracting the associated key objects from each instruction steps, respectively.

\subsection{Evaluations of Consumption Pipeline}\label{sec::app::study_results::consuming}
Fig.~\ref{fig::appenx::consumption_results::tlx} demonstrates the NASA TLX responses of each perceived workloads from participants PC1 - PC12, while using baseline and \sysname~interfaces to perform the designated tasks.
Fig.~\ref{fig::appenx::consumption_results::sus} provides supplementary material of survey results of SUS questionnaires.
To be consistent with the remaining of this paper, we use ``B/L'' to refer to the baseline interface, and ``P2P'' to indicate \sysname, which delivers spatialized and context-aware instruction step.
Finally, Fig.~\ref{fig::consumption::examples}e - h showed examples of how the virtual instructions steps would be anchored on the key objects with the consumption pipeline of \sysname, which are easy to read and would not occlude the consumer's sight while completing the tasks.

\section{Codebook and Themes from Qualitative Data Analysis}\label{app::codebook}
\begin{figure*}[t]
    \centering
    \includegraphics[width=\textwidth]{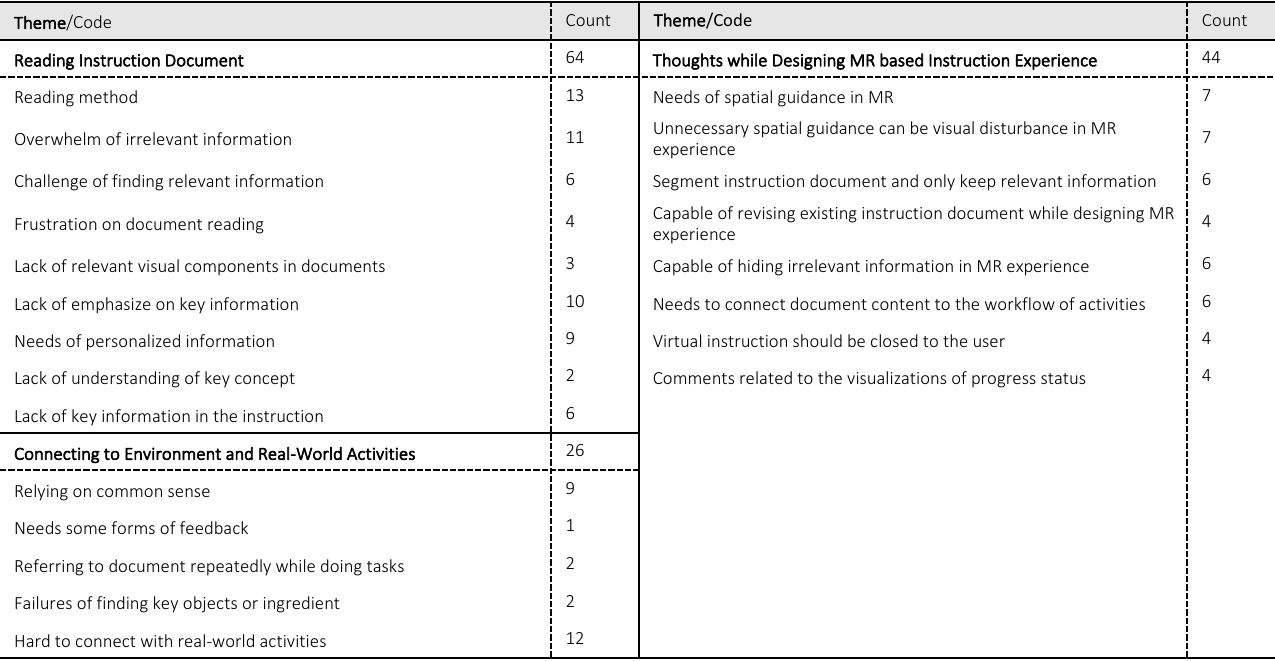}
    \vspace{-0.25in}
    \caption{The codebook that resulted from our qualitative analysis of interview data for preliminary needs-finding study. ``Count'' refers to the number of quote for each theme or code. It is possible that multiple codes are assigned to one quote.}
    \Description{Figure 21 shows a simplified codebook that resulted from our qualitative analysis of interview data for preliminary needs-finding study.}
    \label{fig::appenx::codebook::prelim}
\end{figure*}

\begin{figure*}[t]
    \centering
    \includegraphics[width=\textwidth]{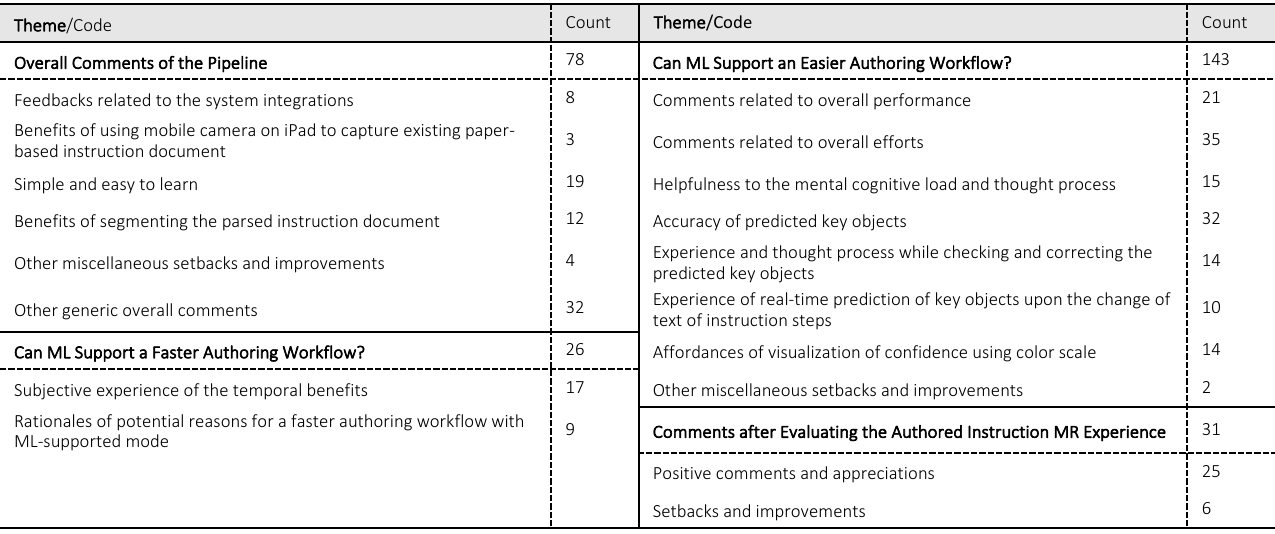}
    \vspace{-0.25in}
    \caption{The codebook that resulted from our qualitative analysis of interview data for authoring pipeline evaluations. ``Count'' refers to the number of quote for each theme or code. It is possible that multiple codes are assigned to one quote.}
    \Description{Figure 22 shows a simplified codebook that resulted from our qualitative analysis of interview data for authoring pipeline evaluations.}
    \label{fig::appenx::codebook::authoring}
\end{figure*}

\begin{figure*}[t]
    \centering
    \includegraphics[width=\textwidth]{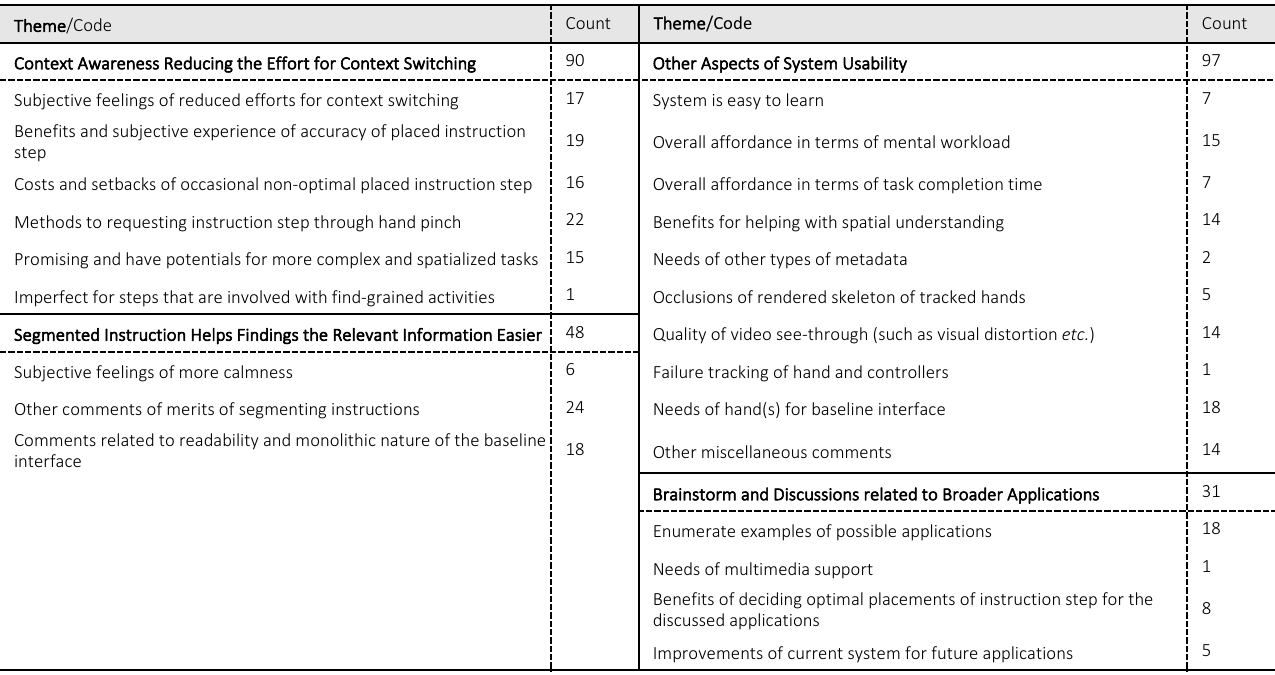}
    \vspace{-0.25in}
    \caption{The codebook that resulted from our qualitative analysis of interview data for consumption pipeline evaluations. ``Count'' refers to the number of quote for each theme or code. It is possible that multiple codes are assigned to one quote.}
    \Description{Figure 23 shows a simplified codebook that resulted from our qualitative analysis of interview data for consumption pipeline evaluations.}
    \label{fig::appenx::codebook::consumption}
\end{figure*}

We used thematic analysis~\cite{Braun2012}, and deductive and inductive coding~\cite{Elo2008} to analyze qualitative data, collected from preliminary needs-finding study (\S\ref{sec::prelim}) and final user study (\S\ref{sec::study}).
As part of supplementary material, we attached the resultant codebook in Fig.~\ref{fig::appenx::codebook::prelim},  Fig.~\ref{fig::appenx::codebook::authoring} and Fig.~\ref{fig::appenx::codebook::consumption}, respectively.
Notably, ``Count'' refers to the number of quote for each theme or code.
It is also possible that multiple codes are assigned to one quote.

\section{Ethical Disclaimer}\label{app::disclaimer}
This work has been approved by the Institutional Review Board (IRB).
All Personal Identifiable Information~(PII), such as the face has been intentionally removed (\eg~being pixelized or blurred) in this manuscript as well as all accompanion videos.
Before each user study, we have obtained participants' consent on video and audio recordings, as well as heterogeneous behavior data collections. 
All participants have consented and acknowledged the data and results presented in this manuscript, to be published and presented publicly for research purposes.
While monetary incentives were not provided, all participants had the opportunity to try out and experience state-of-the-art MR technologies, and to know further about our research.

\end{document}